\documentclass[%
 aip,
 jmp,%
 amsmath,amssymb,
preprint,%
]{revtex4-1}

\usepackage{graphicx}
\usepackage{color}
\usepackage{dcolumn}
\usepackage{bm}

\input xy
\xyoption{all}

\newtheorem{lemma}{Lemma}[section]
\newtheorem{proposition}{Proposition}[section]
\newtheorem{theorem}{Theorem}[section]

\newtheorem{remark}{Remark}[section]

\newtheorem{property}{Property}[section]

\newcommand{\cqfd}{\hfill $\square$}
\begin{document}


\title[] {Generalized Witt and Witt $n-$algebras,  Virasoro algebras and constraints, and KdV equations from $\mathcal{R}(p,q)-$ deformed quantum algebras }

\author{Mahouton Norbert Hounkonnou}
\email{(corresponding author) norbert.hounkonnou@cipma.uac.bj (with copy to hounkonnou@yahoo.fr)}
\affiliation{International Chair in Mathematical Physics and Applications,
	(ICMPA--UNESCO Chair),
	University of Abomey-Calavi,
	072 B.P. 50  Cotonou, Republic of Benin
}%
\author{Fridolin Melong}%
\email{fridomelong@gmail.com}
\affiliation{International Chair in Mathematical Physics and Applications,
	(ICMPA--UNESCO Chair),
	University of Abomey-Calavi,
	072 B.P. 50  Cotonou, Republic of Benin
}
\author{Melanija Mitrovi\'c}
\email{melanija.mitrovic@masfak.ni.ac.rs}
\affiliation{Faculty of Mechanical Engineering, 	
	University of Ni\v s, Serbia
}



\date{\today}

\begin{abstract}
 We perform   generalizations   of  Witt and Virasoro algebras, and derive the corresponding Korteweg-de Vries equations from known $\mathcal{R}(p,q)-$ deformed quantum  algebras previously  introduced in {\it J. Math. Phys.} {\bf 51}, 063518, 
	(2010). Related relevant properties are investigated and discussed. Besides, we construct the $\mathcal{R}(p,q)-$ deformed Witt $n-$ algebra, and determine the Virasoro constraints for a toy model, which play an important role in the study of matrix models. Finally, as matter of illustration,  explicit results are provided for main particular deformed quantum algebras known in the literature.   
\end{abstract}

\keywords{ $\mathcal{R}(p,q)-$ calculus, Witt algebra, Witt $n-$ algebra,  Virasoro algebra, Korteweg-de Vries equation, Virasoro constraints.}

\maketitle


\section{Introduction} 
The Virasoro algebra 
plays an  important
role in physics \cite{S}. 
Kupershmidt \cite{KB} investigated the nature of this algebra \cite{GF}, and discussed its applications in mathematics and physics, namely in  conformal field theory and string theory \cite{BPZ,RY}. Various deformations and generalizations of this algebra were studied in the literature\cite{AS,CJ,Hounkonnou:2015laa}.
For instance,  Hounkonnou et {\it al} generalized Kupershmidt's work \cite{KB}, relatively to  their left-symmetry structure, derived related algebraic and some hydrodynamic properties\cite{Hounkonnou:2015laa}. The realizations of  Witt and Virasoro algebras, and their link with integrable equations were addressed in \cite{HZ}.
 Besides, the link between the Korteweg-de Vries (KdV) equation and the Virasoro algebra was pointed out  by Gervais \cite{G} and Kupershmidt \cite{KB}.

The $q-$analogue of the Lie algebras plays an important role in the study of integrable quantum field theories,  solvable statistical models, string models, and related topics in physics and  mathematics \cite{BL,W}.
Sato \cite{S} investigated the $q-$ version of the deformed Virasoro algebra introduced by Curtright and Zachos \cite{CZ}. New results on the central extension and the operator product expansion were presented and the link  between the $q-$  Virasoro algebra and the Volterra Poisson bracket algebra was discussed.

  The $q-$ deformed KdV equation corresponding to the $q-$ deformed Virasoro algebra was   obtained by Chaichian et {\it al} \cite{CPPa}. This equation can be considered as a lattice system which is a  particular discretization of KdV and a deformation of conformal field theory .

Chaichian et {\it al} \cite{CILPP} also described the $q-$ deformations of the realisations of the conformal algebra depending on the conformal dimension $\Delta.$ The $q-$ deformed Jacobi identity,  $q-$ deformed central extension, $q-$ deformed energy-momentum tensor corresponding to the conformal dimension $\Delta=2,$ and the transformation properties consistent with the $q-$ deformed central extension term  were also considered in their work.
Chaichian and  Presnajder  \cite{CP}   realized the $q-$ deformed  Virasoro algebra using the bosonic annihilation and creation operators of the $q-$ deformed infinite
Heisenberg algebra, and expressed the generators of this new   algebra  as a Sugawara construction. They also presented the  fermonic annihilation and creation operators associated to
the $q-$ deformed infinite Heisenberg superalgebras.

Furthermore, the central extension of the $q-$ deformed Witt algebra and the realization of the $q-$ deformed Virasoro algebra using the $q-$ deformed operator product expansion were studied in\cite{AS}. From their side,
	Wang  et {\it al} \cite{WYWZ}
presented the $q-$ deformed Witt algebra and perfomed their $n-$ algebras. The super $q-$ deformed Virasoro $n-$ algebra for $n$ even  and a toy model for the $q-$ deformed Virasoro constraints were investigated. See also the work by
Nedelina
	and Zabzine on  the $q-$ Virasoro constraints for a toy model \cite{NZ}.

Chakrabarty and  Jagannathan  \cite{CJ} analyzed a $(p,q)$-deformation of the Virasoro algebra with conformal dimension $\Delta$, and defined the 
comultiplication rule for the generating functional for the case $\Delta=0,1.$ The central charge term for the Virasoro algebra associated to the Jagannathan-Srinivasa deformation \cite{JS} was described, and  the corresponding $(p, q)-$ deformed nonlinear equation, also called $(p, q)-$ Korteweg-de Vries equation,  for the case  $\Delta=0$ was derived.

Recently, we investigated  generalizations of $(p,q)-$ deformed Heisenberg algebras,  called $\mathcal{R}(p,q)-$ deformed quantum algebras, where $\mathcal{R}$ is a meromorphic function \cite{HB1}. Furthermore, we  characterized the $\mathcal{R}(p,q)-$ deformed conformal Virasoro algebra,  deduced the $\mathcal{R}(p,q)-$ Korteweg-de Vries equation for a conformal dimension $\Delta=1,$ and discussed the energy-momentum tensor induced by the ${\mathcal R}(p,q)-$ deformed quantum algebras   for the conformal dimension $\Delta=2,$ \cite{HM}.

Before dealing with the main results, let us    fix, in this section, the notations and briefly recall some definitions and known results  useful in the sequel.
	Let $p$ and $q$ be two positive real numbers such that $ 0<q<p\leq1.$ We consider a meromorphic function ${\mathcal R}$ defined on $\mathbb{C}\times\mathbb{C}$ by\cite{HB1}\begin{equation}\label{r10}
	\mathcal{R}(u,v)= \sum_{s,t=-l}^{\infty}r_{st}u^sv^t,
	\end{equation}
	with an eventual isolated singularity at the zero, 
	where $r_{st}$ are complex numbers, $l\in\mathbb{N}\cup\left\lbrace 0\right\rbrace,$ $\mathcal{R}(p^n,q^n)>0,  \forall n\in\mathbb{N},$ and $\mathcal{R}(1,1)=0$ by definition. We denote by $\mathbb{D}_{R}$ the bidisk \begin{eqnarray*}
		\mathbb{D}_{R}&:=&\prod_{j=1}^{2}\mathbb{D}_{R_j}\nonumber\\
		&=&\left\lbrace w=(w_1,w_2)\in\mathbb{C}^2: |w_j|<R_{j} \right\rbrace,
	\end{eqnarray*}
	where $R$ is the convergence radius of the series (\ref{r10}) defined by Hadamard formula as follows:
	\begin{eqnarray*}
		\lim\sup_{s+t \longrightarrow \infty} \sqrt[s+t]{|r_{st}|R^s_1\,R^t_2}=1.
	\end{eqnarray*}
	For the proof and more details see \cite{TN}. Let us also consider $\mathcal{O}(\mathbb{D}_{R})$ the set of holomorphic functions defined on $\mathbb{D}_{R}.$

Define the  $\mathcal{R}(p,q)-$ deformed numbers  \cite{HB}
\begin{equation}\label{rpq number}
[n]_{\mathcal{R}(p,q)}:=\mathcal{R}(p^n,q^n),\quad n\in\mathbb{N}\cup\{0\},
\end{equation}
the
$\mathcal{R}(p,q)-$ deformed factorials
\begin{equation*}\label{s0}
[n]!_{\mathcal{R}(p,q)}:=\left \{
\begin{array}{l}
1\quad\mbox{for}\quad n=0\\
\\
\mathcal{R}(p,q)\cdots\mathcal{R}(p^n,q^n)\quad\mbox{for}\quad n\geq 1,
\end{array}
\right .
\end{equation*}
and the  $\mathcal{R}(p,q)-$ deformed binomial coefficients
\begin{eqnarray*}\label{bc}
\bigg[\begin{array}{c} m  \\ n\end{array} \bigg]_{\mathcal{R}(p,q)} := \frac{[m]!_{\mathcal{R}(p,q)}}{[n]!_{\mathcal{R}(p,q)}[m-n]!_{\mathcal{R}(p,q)}},\quad m,n\in\mathbb{N}\cup\{0\},\quad m\geq n
\end{eqnarray*}
satisfying the relation
\begin{equation*}
\bigg[\begin{array}{c} m  \\ n\end{array} \bigg]_{\mathcal{R}(p,q)}=\bigg[\begin{array}{c} m  \\ m-n\end{array} \bigg]_{\mathcal{R}(p,q)},\quad m,n\in\mathbb{N}\cup\{0\},\quad m\geq n.
\end{equation*}
Consider the following linear operators defined on  $\mathcal{O}(\mathbb{D}_{R}),$ (see \cite{HB1} for more details),
\begin{eqnarray*}
\;Q:\varPsi\longmapsto Q\varPsi(z):&=& \varPsi(qz),\\
\; P:\varPsi\longmapsto P\varPsi(z):&=& \varPsi(pz),\\
\;D_{p,q}:\varPsi\longmapsto D_{p,q}\varPsi(z):&=&\frac{\varPsi(pz)-\varPsi(qz)}{z(p-q)},
\end{eqnarray*}
and the $\mathcal{R}(p,q)-$ derivative 
\begin{equation}\label{r5}
\partial_{\mathcal{R}( p,q)}:=\partial_{p,q}\frac{p-q}{P-Q}\mathcal{R}( P,Q)=\frac{p-q}{p^{P}-q^{Q}}\mathcal{R}(p^{P},q^{Q})\partial_{p,q}.
\end{equation}
The  algebra associated with the $\mathcal{R}(p,q)-$ deformation is a quantum algebra, denoted $\mathcal{A}_{\mathcal{R}(p,q)},$ generated by the set of operators $\{1, A, A^{\dagger}, N\}$ satisfying the following commutation relations:
\begin{eqnarray*}
	&& \label{algN1}
	\quad A A^\dag= [N+1]_{\mathcal {R}(p,q)},\quad\quad\quad A^\dag  A = [N]_{\mathcal {R}(p,q)}.
	\cr&&\left[N,\; A\right] = - A, \qquad\qquad\quad \left[N,\;A^\dag\right] = A^\dag
\end{eqnarray*}
with the realization on  ${\mathcal O}(\mathbb{D}_R)$ given by:
\begin{eqnarray*}\label{algNa}
	A^{\dagger} := z,\qquad A:=\partial_{\mathcal {R}(p,q)}, \qquad N:= z\partial_z,
\end{eqnarray*} 
where $\partial_z:=\frac{\partial}{\partial z}$ is the  derivative on $\mathbb{C}.$

	The Witt algebra ${\mathcal W}$ is the Lie algebra,  which consists of derivations on the Laurent polynomial ring ${\mathbb K}[z,\,z^{-1}],$
given by \cite{IK}:
$${\mathcal W}={\mathbb K}[z,\,z^{-1}]\,{d\over d\,z}.$$ Setting $l_n:=-z^{n+1}\,{d\over d\,z},$ then
$${\mathcal W}=\bigoplus_{n\in\mathbb{Z}}{\mathbb K}\,l_n,$$ and these generators satisfy the commutation relations: $$\big[l_n, l_m\big] = (n-m)\,\,l_{n+m}$$
The Virasoro algebra
$${\mathcal V}ir=\bigoplus_{n\in{\mathbb Z}}{\mathbb K}L_n \oplus {\mathbb K}\,C$$ is the Lie algebra which satisfies the following commutation relations\cite{IK}: 
\begin{eqnarray}\label{va}
\left[L_n, L_m \right]=(n-m)L_{n+m} + \frac{1}{12}(n^3-n)\delta_{n+m,o}\,C,
\end{eqnarray}
\begin{eqnarray*}
\left[{\mathcal V}ir, C\right]=\{0\},
\end{eqnarray*}
where $\delta_{i,j}$ denotes the Kronecker delta and $C$ the central charge.

Let us finish this series of remindings by pointing out   the connection between the
Virasoro algebra and the Koterweg-de Vries (KdV) equation. For that, we use the relation (\ref{va}). In order
to give the realization of the algebra $\mathcal{V}$ in terms of the currents which satisfy the Korteweg-de Vries (KdV) equation, Chaichian {\it et al} define the current  as follows \cite{CPPa}:
\begin{eqnarray*}
	u(x):= {6\over c}\,\displaystyle\sum_{-\infty}^{\infty}l_n\,e^{-i\,n\,x}-{1\over 4}.
\end{eqnarray*}
The commutation relation (\ref{va})
yields the relation \cite{G}:
\begin{eqnarray*}
	\Big[u(x),u(y)\Big]&=& P\,\delta(x-y)\nonumber\\&=&\Big(\partial_{xxx}+ 2\,u\partial_{x} + u_x\Big)\,\delta(x-y),
\end{eqnarray*}
where $P$ can be considered as the Hamiltonian operator \cite{GF}  for which the Schouten bracket vanishes \cite{FF}, and hence it can be used to construct Hamiltonian systems. Indeed, defining  the so-called second Hamiltonian as :
\begin{eqnarray*}
	H={1\over 2}\int u^2(x)\,d\,x
\end{eqnarray*}
leads to the Hamiltonian equation
\begin{eqnarray*}
	u_t=P\,grad\,H= u_{xxx} + 6\,u\,u_{x}
\end{eqnarray*}
which is the Korteweg-de Vries equation.

In  this paper, we construct the  Witt algebra, its $n-$ version, the Virasoro algebra and its constraints for a toy model, and the Korteweg-de Vries equation induced by the $\mathcal{R}(p,q)-$ deformed quantum algebras \cite{HB}. We derive particular cases associated to quantum algebras developed in the literature, and discuss  their main properties.

This paper is organized as follows.
In section 2, we investigate the construction of the  $\mathcal{R}(p,q)-$ deformed Witt algebra and derive its main properties.
Section 3 is dedicated to study  the $\mathcal{R}(p,q)-$ deformed Virasoro algebra and associated
 Korteweg-de Vries (KdV) equation.
 We deduct   particular cases corresponding to deformed quantum algebras known in the literature.
In section 4, we investigate the related Witt $n-$ algebra and a toy model, and 
 deduce relevant specific cases. 
We end with the concluding remarks in section 5.
\section{$\mathcal{R}(p,q)-$ deformed Witt algebra}
In this section,  we study  the $\mathcal{R}(p,q)-$ deformed Witt algebra by using 
the $\mathcal{R}(p,q)-$ derivative (\ref{r5}) 
\begin{equation}\label{Rpqd}
\partial_{\mathcal{R}(p,q)}\varphi(z):= {1\over z}\,[z\partial_z]_{\mathcal{R}(p,q)}\varphi(z)
\end{equation}
which generalizes particular derivatives known in the literature as follows:
\begin{itemize}
	\item[(i)]  $q-$ Heine derivative \cite{Heine}
	\begin{equation*}
	\mathcal{R}(1,q)=1
	\quad \mbox{and} \quad
	\partial_{q}\varphi(z)=\frac{1}{z}\,[z\partial_z]_q\,\varphi(z);
	\end{equation*} 
	\item[(ii)]  $q-$ Quesne  derivative \cite{QPT}
	\begin{equation*}
	\mathcal{R}(1,q)={1-q^{-1} \over q-1}
	\quad \mbox{and}\quad
	\partial_{q}\varphi(z)=\frac{1}{z}\,[z\partial_z]_{q}\,\varphi(z);
	\end{equation*}
	\item[(iii)] $(p,q)-$ Jagannathan-Srinivasa derivative \cite{JS}
	\begin{equation*}
	\mathcal{R}(p,q)=1\quad\mbox{and}\quad \partial_{p,q}\varphi(z)=\frac{1}{z}\,[z\partial_z]_{p,q}\,\varphi(z);
	\end{equation*}
	\item[(iv)] $(p^{-1},q)-$ Chakrabarty - Jagannathan derivative \cite{Chakrabarti&Jagan}
	\begin{eqnarray*}
		\mathcal{R}(p,q)=1
		\quad \mbox{and} \quad
		\partial_{p^{-1},q}\varphi(z)=\frac{1}{z}\,[z\partial_z]_{p^{-1},q}\,\varphi(z);
	\end{eqnarray*}
	\item[(v)] Hounkonnou-Ngompe  generalization of $q-$ Quesne derivative \cite{HN}
	\begin{eqnarray*}
		\mathcal{R}(p,q)={p\,q-1 \over (q-p^{-1})q}
		\quad \mbox{and} \quad 
		\partial_{p,q}\varphi(z)=\frac{1}{z}\,[z\partial_z]_{p,q}\,\varphi(z);
	\end{eqnarray*}
\end{itemize}

Let us introduce the model deformation structure functions  $\epsilon_i, i\in\{1,2\},$ depending on the deformation parameters $p$ and $q,$  which allow to redefine the ${\mathcal R}(p, q)-$numbers (\ref{rpq number}) as
\begin{eqnarray}\label{rpqn}
[n]_{\mathcal{R}(p,q)}:= {\epsilon^n_1- \epsilon^n_2\over \epsilon_1-\epsilon_2},\quad\quad \epsilon_1\ne \epsilon_2
\end{eqnarray}
from which known  particular cases can be   deduced,   namely:
\begin{itemize}
	\item [(i)]$q-$ Arick-Coon-Kuryskin  deformation \cite{AC}
	\begin{equation*}
	\epsilon_1=1, \quad \epsilon_2=q \quad \mbox{and} \quad
	[n]_q ={1-q^n\over 1-q};
	\end{equation*}
	\item[(ii)]  $q-$ Quesne deformation \cite{QPT}
	\begin{equation*}
	\epsilon_1=1, \quad \epsilon_2=q^{-1} \quad \mbox{and} \quad
	[n]_q ={1-q^{-n}\over q-1};
	\end{equation*}
	\item [(iii)]$(p,q)-$ Jagannathan-Srinivasa deformation \cite{JS}
	\begin{equation}\label{JSn}
	\epsilon_1=p, \quad \epsilon_2=q \quad \mbox{and} \quad
	[n]_{p,q} ={p^n-q^n\over p-q};
	\end{equation}
	\item [(iv)]$(p^{-1},q)-$ Chakrabarty -Jagannathan deformation \cite{CJ}
	\begin{equation*}
	\epsilon_1=p^{-1}, \quad \epsilon_2=q \quad \mbox{and} \quad
	[n]_{p^{-1},q} ={p^{-n}-q^{n}\over p^{-1}-q};
	\end{equation*}
	\item [(v)] Hounkonnou-Ngompe  generalization of $q-$ Quesne deformation \cite{HN} 
	\begin{equation*}
	\epsilon_1=p, \quad \epsilon_2=q^{-1} \quad \mbox{and} \quad
	[n]_{p,q} ={p^{n}-q^{-n}\over q-p^{-1}};
	\end{equation*}
\end{itemize}

This $\mathcal{R}(p,q)$- deformed Witt algebra is spanned by  the  generators $e^{\mathcal{R}(p,q)}_n$ acting on a  holomorphic function $\varphi$ as
\begin{eqnarray*}
	e^{\mathcal{R}(p,q)}_n\,\varphi(z) :=z^{n+1}\,\partial_{\mathcal{R}(p,q)}\,\varphi(z),
\end{eqnarray*}
or, equivalently, 
from the relation (\ref{Rpqd}),
\begin{eqnarray*}
e^{\mathcal{R}(p,q)}_n\,\varphi(z) 
=\big[z\partial_z-n\big]_{\mathcal{R}(p,q)}\,z^n\,\varphi(z).
\end{eqnarray*}
\begin{lemma}
	The  generators $e^{\mathcal{R}(p,q)}_n$  satisfy the commutation relation:
	
	\begin{eqnarray*}\label{cs}
	\left[e^{\mathcal{R}(p,q)}_n,e^{\mathcal{R}(p,q)}_m \right]_{s, t}={1\over \epsilon_1-\epsilon_2}\Big(\epsilon^{z\partial_z}_1\big(s\,\epsilon^{-n}_1-t\,\epsilon^{-m}_1\big)-\epsilon^{z\partial_z}_2\big(s\,\epsilon^{-n}_2-t\,\epsilon^{-m}_2\big)\Big)e^{\mathcal{R}(p,q)}_{n+m},
	\end{eqnarray*}
	where $$ \big[A,\,B\big]_{u,v}= u\,AB-v\,BA,$$
	$u$ and $v$ are arbitrary real (or complex) numbers.
  For  $s=1$ and $t=\epsilon^{m-n}_2,$ we obtain:
	\begin{eqnarray*}
	\Big[e^{\mathcal{R}(p,q)}_n,e^{\mathcal{R}(p,q)}_m\Big]_{1,\epsilon^{m-n}_2}=\,[m-n]_{\mathcal{R}(p,q)}\,\epsilon^{{z\partial_z}-m}_1\,e^{\mathcal{R}(p,q)}_{n+m}.
	\end{eqnarray*}
\end{lemma}
{\it Proof.}  It stems from a straightforward computation.
 \cqfd

This suggests to  define new  $\mathcal{R}(p,q)-$ deformed generators 
\begin{eqnarray*}\label{Rrgo}
l^{\mathcal{R}(p,q)}_n\varphi(z) &:=&\epsilon^{-z\partial_z}_1\,\big[z\partial_z-n\big]_{\mathcal{R}(p,q)}\,z^n\,\varphi(z)
\end{eqnarray*}
whose the commutator obeys the conventional Witt algebra structure given by the next Theorem \ref{convWitt}.
\begin{theorem}\label{convWitt}
	\begin{eqnarray}\label{RWa}
	\Big[l^{\mathcal{R}(p,q)}_n,l^{\mathcal{R}(p,q)}_m\Big]_{\epsilon^{m-n}_1,\epsilon^{m-n}_2} = [m-n]_{\mathcal{R}(p,q)}\,l^{\mathcal{R}(p,q)}_{n+m}
	\end{eqnarray}
	and
	\begin{eqnarray}\label{Rwb}
	\Big[l^{\mathcal{R}(p,q)}_n,l^{\mathcal{R}(p,q)}_m\Big]=[m-n]_{\mathcal{R}(p,q)}\,\epsilon^{-{z\partial_z}+n}_1\epsilon^{{z\partial_z}-m}_2\,l^{\mathcal{R}(p,q)}_{n+m}.
	\end{eqnarray}
\end{theorem}

		The merit of the relations (\ref{RWa}) and (\ref{Rwb}) consists of obtaining the ${\mathcal R}(p,q)-$ deformed $su(1,1)$ subalgebra:
		\begin{eqnarray*}
		\,\Big[l^{\mathcal{R}(p,q)}_0,l^{\mathcal{R}(p,q)}_1\Big]_{\epsilon_1,\epsilon_2} &=& \,l^{\mathcal{R}(p,q)}_{1},\nonumber\\
		\,
		\Big[l^{\mathcal{R}(p,q)}_{-1},l^{\mathcal{R}(p,q)}_0\Big]_{\epsilon_1,\epsilon_2} &=&\,l^{\mathcal{R}(p,q)}_{-1},
		\end{eqnarray*}
		and
		\begin{eqnarray*}
		\Big[l^{\mathcal{R}(p,q)}_{-1},l^{\mathcal{R}(p,q)}_{1}\Big]=[2]_{\mathcal{R}(p,q)}\,\epsilon^{-{z\partial_z}-1}_1\epsilon^{{z\partial_z}-1}_2\,l^{\mathcal{R}(p,q)}_{0}.
		\end{eqnarray*}

Note that:
\begin{itemize}
	\item
 taking $\mathcal{R}(p,q)=1$ and  $\epsilon_1=p^{-1},$ we recover the $(p,q)-$ deformed Witt algebra displayed in \cite{HM} with the generators $l^{p,q}_n$ acting as follows:
	\begin{eqnarray*}
		l^{p,q}_n\varphi(z) =p^{z\partial_z}\,\big[z\partial_z-n\big]_{p,q}\,z^n\,\varphi(z)
	\end{eqnarray*}
	and satisfying the commutation relations
	\begin{eqnarray*}
		\,\Big[l^{p,q}_n,l^{p,q}_m\Big]_{p^{n-m},q^{m-n}} &=& [m-n]_{p,q}\,l^{p,q}_{n+m},\\
		\,\Big[l^{p,q}_n,l^{p,q}_m\Big]&=&[m-n]_{p,q}\,p^{{z\partial_z}-n}\,q^{{z\partial_z}-m}\,l^{p,q}_{n+m},
	\end{eqnarray*}
	with \begin{eqnarray}\label{pqn}
	[x]_{p,q}={q^x-p^{-x}\over q-p^{-1}};
	\end{eqnarray}
	\item the $q-$ deformed Witt algebra given in \cite{CPPa} can be obtained by putting $\epsilon_1=q$ and $\epsilon_2=q^{-1},$ with the generators $l^{q}_n:$
	\begin{eqnarray*}
		l^{q}_n\varphi(z) =\,q^{-z\partial_z}\,\big[z\partial_z-n\big]_{q}\,z^n\,\varphi(z)
	\end{eqnarray*}
obeying the law
	\begin{eqnarray*}
		\,\Big[l^{q}_n,l^{q}_m\Big]_{q^{m-n},q^{n-m}} &=& [m-n]_{q}\,l^{q}_{n+m},\\
		\,\Big[l^{q}_n,l^{q}_m\Big]&=&[m-n]_{q}\,q^{-{z\partial_z}+n}\,q^{-{z\partial_z}+m}\,l^{q}_{n+m},
	\end{eqnarray*}
	and
	\begin{eqnarray*}
		[x]_q={q^x-q^{-x}\over q-q^{-1}}.
	\end{eqnarray*}
\end{itemize}
\begin{remark}  We can  readily deduct the Witt algebras for classes of known deformations as follows:
	\begin{itemize}
		\item[(i)] Taking $\mathcal{R}(1,q)=1,$ which implies $\epsilon_1=1$ and $\epsilon_2=q,$ we obtain the $q-$ deformed generators:
		\begin{eqnarray*}
			l^{q}_n\varphi(z) =\,\big[z\partial_z-n\big]_{q}\,z^n\,\varphi(z)
		\end{eqnarray*} 
		satisfying the relations:
		\begin{eqnarray*}
			\,\Big[l^{q}_n,l^{q}_m\Big]_{1,q^{m-n}} &=& [m-n]_{q}\,l^{q}_{n+m},\\
			\,\Big[l^{q}_n,l^{q}_m\Big]&=&[m-n]_{q}\,q^{{z\partial_z}-m}\,l^{q}_{n+m}.
		\end{eqnarray*}
		\item[(ii)] The Witt algebra  associated to the {\bf Jagannathan - Srinivasa} deformation \cite{JS} is obtained with $\epsilon_1=p,$  $\epsilon_2=q,$  and
		\begin{eqnarray*}
			l^{p,q}_n\varphi(z) =\,p^{-z\partial_z}\,\big[z\partial_z-n\big]_{p,q}\,z^n\,\varphi(z),
		\end{eqnarray*}
		satisfying
		\begin{eqnarray*}
			\,\Big[l^{p,q}_n,l^{p,q}_m\Big]_{p^{m-n} , q^{m-n}} &=& [m-n]_{p,q}\,l^{p,q}_{n+m},\\
			\,\Big[l^{p,q}_n,l^{p,q}_m\Big]&=&[m-n]_{p,q}\,p^{-{z\partial_z}+n}\,q^{{z\partial_z}-m}\,l^{p,q}_{n+m}.
		\end{eqnarray*}
		\item [(iii)]For $\epsilon_1=p^{-1}$ and $\epsilon_2=q,$ we get the Witt algebra associated to the {\bf Chakrabarty-Jagannathan} deformation \cite{Chakrabarti&Jagan}, with
		\begin{eqnarray*}
			l^{p^{-1},q}_n\varphi(z) =\,p^{z\partial_z}\,\big[z\partial_z-n\big]_{p^{-1},q}\,z^n\,\varphi(z),
		\end{eqnarray*}
		and 
		\begin{eqnarray*}
			\,\Big[l^{p^{-1},q}_n,l^{p^{-1},q}_m\Big]_{p^{n-m},q^{m-n}} &=& [m-n]_{p^{-1},q}\,l^{p^{-1},q}_{n+m},\\
			\,\Big[l^{p^{-1},q}_n,l^{p^{-1},q)}_m\Big]&=&[m-n]_{p^{-1},q}\,p^{{z\partial_z}-n}\,q^{{z\partial_z}-m}\,l^{p^{-1},q}_{n+m}.
		\end{eqnarray*}
	Moreover, the $(p^{-1},q)-$ deformed $su(1,1)$ subalgebra is here furnished by:
		\begin{eqnarray*}
		\,\Big[l^{p^{-1},q}_0,l^{p^{-1},q}_1\Big]_{p^{-1},q} &=& \,l^{p^{-1},q}_{1},\nonumber\\
		\,
		\Big[l^{p^{-1},q}_{-1},l^{p^{-1},q}_0\Big]_{p^{-1},q} &=&\,l^{p^{-1},q}_{-1},\nonumber\\
		\,\Big[l^{p^{-1},q}_{-1},l^{p^{-1},q}_{1}\Big]&=&[2]_{p^{-1},q}\,p^{z\partial_z+1}\,q^{{z\partial_z}-1}\,l^{p^{-1},q}_{0}.
		\end{eqnarray*}
		\item[(iv)] The Witt algebra corresponding to {\bf Hounkonnou-Ngompe } generalization of  $q-$ Quesne  deformation \cite{HN} is built by taking $\epsilon_1=p,$ $\epsilon_2=q^{-1},$ 	\begin{eqnarray*}
			l^{p,q}_n\varphi(z) =\,p^{-z\partial_z-1}\,q\,\big[z\partial_z-n\big]_{p,q}\,z^n\,\varphi(z),
		\end{eqnarray*}
		with the relations:
		\begin{eqnarray*}
			\,\Big[\,l^{p,q}_n,\,l^{p,q}_m\Big]_{p^{m-n},q^{n-m}} &=&qp^{-1}\, [m-n]_{p,q}\,l^{p,q}_{n+m},\\
			\,\Big[\,l^{p,q}_n,\,l^{p,q}_m\Big]&=&qp^{-1}\, [m-n]_{p,q}\,p^{-{z\partial_z}+n}\,q^{-{z\partial_z}+m}\,l^{p,q}_{n+m}.
		\end{eqnarray*}
	Furthermore, its $su(1,1)$ subalgebra is realized as follows:\begin{eqnarray*}
			\,\Big[\,l^{p,q}_0,\,l^{p,q}_1\Big]_{p,q^{-1}} &=&qp^{-1}\,l^{p,q}_{1},\\
			\,\Big[\,l^{p,q}_{-1},\,l^{p,q}_0\Big]_{p,q^{-1}} &=&qp^{-1}\, l^{p,q}_{-1},\\
			\,\Big[\,l^{p,q}_{-1},\,l^{p,q}_1\Big]&=&qp^{-1}\, [2]_{p,q}\,p^{-{z\partial_z}-1}\,q^{-{z\partial_z}+1}\,l^{p,q}_{0}.
	\end{eqnarray*}
	\end{itemize}
\end{remark}
The $\mathcal{R}(p,q)$-deformed Witt algebra (\ref{RWa}) satisfies the Jacobi identity:

\begin{eqnarray}\label{RJ}
\displaystyle
\sum_{(i,j,l)\in\mathcal{C}(n,m,k)}{1\over (\epsilon_1\,\epsilon_2)^{l}}\frac{[2i]_{\mathcal{R}(p,q)}}{[i]_{\mathcal{R}(p,q)}}\big[l^{\mathcal{R}(p,q)}_i, \big[l^{\mathcal{R}(p,q)}_j , l^{\mathcal{R}(p,q)}_l\big]_{x_{jl},y_{jl}}\big]_{x_{i(j+l)},y_{i(j+l)}} =0,
\end{eqnarray}
where  $n$, $m$ and $k$ are natural numbers, and  $\mathcal{C}(n,m,k)$ refers  to the
cyclic permutation of $(n,m,k)$.
\begin{remark} Particular cases of  
	the $\mathcal{R}(p,q)$-deformed Jacobi identity (\ref{RJ}) are deduced as follows:
	\begin{enumerate} 
		\item[(i)] For the $q-$ deformed  algebra in \cite{CILPP}:
		\begin{eqnarray*}
			\displaystyle
			\sum_{(i,j,l)\in\mathcal{C}(n,m,k)}\frac{[2i]_{q}}{[i]_{q}}\big[l^{q}_i, \big[l^{q}_j , l^{q}_l\big]_{x_{jl},y_{jl}}\big]_{x_{i(j+l)},y_{i(j+l)}} =0.
		\end{eqnarray*}
		\item[(ii)] For the  {\bf Arick-Coon} $q-$ deformation \cite{AC}:
		\begin{eqnarray*}
			\displaystyle
			\sum_{(i,j,l)\in\mathcal{C}(n,m,k)}{1\over q^{l}}\frac{[2i]_{q}}{[i]_{q}}\big[l^{q}_i, \big[l^{q}_j , l^{q}_l\big]_{x_{jl},y_{jl}}\big]_{x_{i(j+l)},y_{i(j+l)}} =0.
		\end{eqnarray*}
		\item[(iii)] For   the $(p,q)-$ deformed  algebra in  \cite{CJ}:
		\begin{eqnarray*}
			\displaystyle
			\sum_{(i,j,l)\in\mathcal{C}(n,m,k)}\Big({p\over q}\Big)^{l}\frac{[2i]_{p,q}}{[i]_{p,q}}\big[l^{p,q}_i, \big[l^{p,q}_j , l^{p,q}_l\big]_{x_{jl},y_{jl}}\big]_{x_{i(j+l)},y_{i(j+l)}} =0.
		\end{eqnarray*}
		\item[(iv)] For the \textbf{Jagannathan-Srinivasa} deformation \cite{JS}: 
		\begin{eqnarray*}
			\displaystyle
			\sum_{(i,j,l)\in\mathcal{C}(n,m,k)}{1\over (pq)^{l}}\frac{[2i]_{p,q}}{[i]_{p,q}}\big[l^{p,q}_i, \big[l^{p,q}_j , l^{p,q}_l\big]_{x_{jl},y_{jl}}\big]_{x_{i(j+l)},y_{i(j+l)}} =0.
		\end{eqnarray*}
		\item[(v)] For the \textbf{Chakrabarty} and \textbf{Jagannathan} deformation \cite{Chakrabarti&Jagan}:
		\begin{eqnarray*}
			\displaystyle
			\sum_{(i,j,l)\in\mathcal{C}(n,m,k)}{1\over ({q\over p})^{l}}\frac{[2i]_{p^{-1},q}}{[i]_{p^{-1},q}}\big[l^{p^{-1},q}_i, \big[l^{p^{-1},q}_j , l^{p^{-1},q}_l\big]_{x_{jl},y_{jl}}\big]_{x_{i(j+l)},y_{i(j+l)}} =0.
		\end{eqnarray*}
		\item[(vi)] For the \textbf{Hounkonnou-Ngompe} generalization of $q-$ Quesne deformation  \cite{HN}, corresponding to  $\epsilon_1=p$ and $\epsilon_2=q^{-1}$:
		\begin{eqnarray*}
			\displaystyle
			\sum_{(i,j,l)\in\mathcal{C}(n,m,k)}{1\over ({p\over q})^{l}}\frac{[2i]_{p,q}}{[i]_{p,q}}\big[l^{p,q}_i, \big[l^{p,q}_j , l^{p,q}_l\big]_{x_{jl},y_{jl}}\big]_{x_{i(j+l)},y_{i(j+l)}} =0.
		\end{eqnarray*}
	\end{enumerate}
\end{remark}
\subsection{Some properties of the $\mathcal{R}(p,q)$-deformed Witt algebra}
This section is devoted to a list of  some remarkable identities pertaining to  the $\mathcal{R}(p,q)-$ deformed Witt algebra. 
	For $n\neq m$, the $\mathcal{R}(p,q)-$deformed Witt generator product yields
	\begin{equation*}\label{S12}
	l^{\mathcal{R}(p,q)}_n . l^{\mathcal{R}(p,q)}_m = \mathcal{D}^{\mathcal{R}(p,q)}_{n}\,l^{\mathcal{R}(p,q)}_{n+m},
	\end{equation*} 
	where 
	\begin{equation*}\label{S12a}
	\mathcal{D}^{\mathcal{R}(p,q)}_{n}:=\epsilon^{-z\partial_z+n}_1\,[z\partial_z-n]_{\mathcal{R}(p,q)} .
	\end{equation*}
	The $\mathcal{R}(p,q)-$ deformed Witt algebra 
equipped with the natural operator  product "." is a nonassociative algebra with associator given by the relation :
	\begin{small}
		\begin{eqnarray*}\label{las}
			\Big(l^{\mathcal{R}(p,q)}_{n},l^{\mathcal{R}(p,q)}_{m},l^{\mathcal{R}(p,q)}_{k}\Big)&:=&l^{\mathcal{R}(p,q)}_{n}.\big(l^{\mathcal{R}(p,q)}_{m}.l^{\mathcal{R}(p,q)}_{k}\big)-\big(l^{\mathcal{R}(p,q)}_{n}.l^{\mathcal{R}(p,q)}_{m}\big).l^{\mathcal{R}(p,q)}_{k}\cr
			&=&{\mathcal R^{n}_{km}} 
			- {\mathcal R}^{n}_{k(m+n)},
		\end{eqnarray*}
	\end{small}
where \begin{eqnarray*}
	{\mathcal R^{i}_{lj}}=\epsilon^{-2z\partial_z+i+j}_1[z\partial_z-i]_{\mathcal{R}(p,q)}[z\partial_z-j]_{\mathcal{R}(p,q)}l^{\mathcal{R}(p,q)}_{i+j+l}.
\end{eqnarray*}
	The case $n=0$ leads to an associative algebra as required with the trivial associator:
	\begin{eqnarray*}
		\Big(l^{\mathcal{R}(p,q)}_{0},l^{\mathcal{R}(p,q)}_{m},l^{\mathcal{R}(p,q)}_{k}\Big)=0,
	\end{eqnarray*}
	for all $m$ and $k.$
\begin{property}
	\begin{eqnarray*}
		&&\Big[l^{\mathcal{R}(p,q)}_n,l^{\mathcal{R}(p,q)}_m\Big]=-\Big[l^{\mathcal{R}(p,q)}_m,l^{\mathcal{R}(p,q)}_n\Big].
	\end{eqnarray*}
\end{property}
	\begin{property}
\begin{eqnarray*}
&&\displaystyle\sum_{(i,j,l)
		\in\mathcal {C}(n,m,k)}
	\Big(l^{\mathcal{R}(p,q)}_{i},l^{\mathcal{R}(p,q)}_{j},l^{\mathcal{R}(p,q)}_{l}\Big)-
	\sum_{(i,j,l)\in{\mathcal C}(n,k,m)}\,\Big(l^{\mathcal{R}(p,q)}_{i},l^{\mathcal{R}(p,q)}_{j},l^{\mathcal{R}(p,q)}_{l}\Big)\cr&&\qquad\qquad\qquad\qquad\qquad\qquad\qquad\qquad\qquad\qquad\qquad=\tau^{n}_{mk} + \tau^{k}_{nm} + \tau^{m}_{kn},
	\end{eqnarray*}
	where $\mathcal{C}(n,m,k)$ denotes the
	cyclic permutation of $(n,m,k)$ and 
	\begin{small}
	\begin{eqnarray*}
		\tau^{j}_{il}=\epsilon^{-2z\partial_z+i+j}_1[z\partial_z-i-j]_{\mathcal{R}(p,q)}\Big(\epsilon^{i}_1[z\partial_z-i]_{\mathcal{R}(p,q)}-\epsilon^{j}_1[z\partial_z-j]_{\mathcal{R}(p,q)}\Big)l^{\mathcal{R}(p,q)}_{i+j+l}.
	\end{eqnarray*}
\end{small}
\end{property}
\begin{property}
	\begin{small}
		\begin{eqnarray*}
			\Big[l^{\mathcal{R}(p,q)}_n , l^{\mathcal{R}(p,q)}_m . l^{\mathcal{R}(p,q)}_k \Big] &=& l^{\mathcal{R}(p,q)}_m . \Big[l^{\mathcal{R}(p,q)}_n, l^{\mathcal{R}(p,q)}_k \Big] + \Big[ l^{\mathcal{R}(p,q)}_n, l^{\mathcal{R}(p,q)}_m\Big]. l^{\mathcal{R}(p,q)}_k\cr&+& {\mathcal R}^{m}_{k(m+n)}-{\mathcal R}^{n}_{k(m+n)}+ {\mathcal R}^{m}_{nk}-{\mathcal R}^{m}_{n(m+k)}.
		\end{eqnarray*}
	\end{small}
The ${\mathcal R}(p,q)-$ derivation (Leibniz rule) should be satisfied if 
\begin{eqnarray*}
{\mathcal R}^{m}_{k(m+n)}-{\mathcal R}^{n}_{k(m+n)}= {\mathcal R}^{m}_{n(m+k)}-{\mathcal R}^{m}_{nk}.
\end{eqnarray*}
\end{property}
\begin{property}
	\begin{small}
		\begin{eqnarray*}
			&&\Big(l^{\mathcal{R}(p,q)}_n, l^{\mathcal{R}(p,q)}_m,l^{\mathcal{R}(p,q)}_k\Big)= \Big(l^{\mathcal{R}(p,q)}_m,l^{\mathcal{R}(p,q)}_n ,l^{\mathcal{R}(p,q)}_k\Big) + \tau^{n}_{mk}.
		\end{eqnarray*}
	\end{small}
Therefore,  the ${\mathcal R}(p,q)-$ left symmetry property would require $\tau^{n}_{mk}=0.$ 
\end{property}
\begin{property}
	The Nambu $3-$ bracket  defined by:
		\begin{eqnarray*}
			&&\Big[l^{\mathcal{R}(p,q)}_n, l^{\mathcal{R}(p,q)}_m,l^{\mathcal{R}(p,q)}_k\Big]:=l^{\mathcal{R}(p,q)}_n\Big[ l^{\mathcal{R}(p,q)}_m,l^{\mathcal{R}(p,q)}_k\Big]+l^{\mathcal{R}(p,q)}_m\Big[ l^{\mathcal{R}(p,q)}_k,l^{\mathcal{R}(p,q)}_n\Big]\cr&&\qquad\qquad\qquad\qquad\qquad\qquad\qquad\qquad\qquad\qquad\qquad\qquad + l^{\mathcal{R}(p,q)}_k\Big[ l^{\mathcal{R}(p,q)}_n,l^{\mathcal{R}(p,q)}_m\Big],
	\end{eqnarray*}
 yields
	\begin{small}
		\begin{eqnarray*}\label{tary}
			&&\Big[l^{\mathcal{R}(p,q)}_n, l^{\mathcal{R}(p,q)}_m,l^{\mathcal{R}(p,q)}_k\Big]
			= {\mathcal R^{n}_{km}}+ {\mathcal R^{m}_{nk}} + {\mathcal R^{k}_{mn}}.
		\end{eqnarray*}
	\end{small}
\end{property}
\begin{property}
	\begin{small}
		\begin{eqnarray*}
			&&	\Big[\Big[l^{\mathcal{R}(p,q)}_n, l^{\mathcal{R}(p,q)}_m, l^{\mathcal{R}(p,q)}_k. l^{\mathcal{R}(p,q)}_s\Big]\Big]-\Big[\Big[l^{\mathcal{R}(p,q)}_n, l^{\mathcal{R}(p,q)}_m, l^{\mathcal{R}(p,q)}_k\Big]\Big]. l^{\mathcal{R}(p,q)}_s\cr&&\qquad\qquad\qquad\qquad\qquad\qquad\qquad\qquad -l^{\mathcal{R}(p,q)}_k. \Big[\Big[l^{\mathcal{R}(p,q)}_n, l^{\mathcal{R}(p,q)}_m, l^{\mathcal{R}(p,q)}_s\Big]\Big]= \mathcal{D}^{\mathcal{R}(p,q)}_{n+m+k}\,\tau^{m}_{n\,k}.
		\end{eqnarray*}
	\end{small}
\end{property}
\begin{proposition}
	The following identities, where the left and right multiplication operators are defined, respectively,  by:
		\begin{equation*}
			{\bf L}_{a}(b):= a.b\quad\mbox{and}\quad {\bf R}_{a}(b):
			=  b. a,
	\end{equation*}
 hold:
	\begin{enumerate}
		\item[(i)]
		\begin{small}
			\begin{eqnarray*}
				&&\Big[{\bf L}_{l^{\mathcal{R}(p,q)}_n}, {\bf L}_{l^{\mathcal{R}(p,q)}_m}\Big]l^{\mathcal{R}(p,q)}_{k}={\bf L}_{[l^{\mathcal{R}(p,q)}_n,l^{\mathcal{R}(p,q)}_m]} + \tau^{n}_{mk},\quad \forall k, 
			\end{eqnarray*}
		\end{small}
		\item [(ii)]
		\begin{small}
			\begin{eqnarray*}
				&&{\Big[{\bf L}_{l^{\mathcal{R}(p,q)}_n},\, {\bf R}_{l^{\mathcal{R}(p,q)}_m}\Big]}l^{\mathcal{R}(p,q)}_{k}=l^{\mathcal{R}(p,q)}_n . \Big[l^{\mathcal{R}(p,q)}_{k}, l^{\mathcal{R}(p,q)}_m\Big]+ {\mathcal R^{m}_{kn}} - {\mathcal R^{n}_{m(n+k)}}.	
			\end{eqnarray*}
		\end{small}
		\item [(iii)]
		\begin{small}
			\begin{eqnarray*}
				{\Big[{\bf R}_{l^{\mathcal{R}(p,q)}_n}, {l^{\mathcal{R}(p,q)}_m}\Big]}l^{\mathcal{R}(p,q)}_{k}=l^{\mathcal{R}(p,q)}_m .\Big[l^{\mathcal{R}(p,q)}_n, l^{\mathcal{R}(p,q)}_{k}\Big]- {\mathcal R^{m}_{kn}} + {\mathcal R^{m}_{k(m+n)}}.
			\end{eqnarray*}
		\end{small}
		\item [(iv)]
		\begin{small}
			\begin{eqnarray*}
				&&\Big[{\bf R}_{l^{\mathcal{R}(p,q)}_n}{\bf R}_{l^{\mathcal{R}(p,q)}_m}+{\bf R}_{l^{\mathcal{R}(p,q)}_n l^{\mathcal{R}(p,q)}_m}\Big]l^{\mathcal{R}(p,q)}_{k}=l^{\mathcal{R}(p,q)}_{k}. \Big[{\bf R}_{l^{\mathcal{R}(p,q)}_n}(l^{\mathcal{R}(p,q)}_m)+ {\bf R}_{l^{\mathcal{R}(p,q)}_m}(l^{\mathcal{R}(p,q)}_n)\Big]\cr&&\qquad\qquad\qquad\qquad\qquad\qquad\qquad\qquad\qquad\qquad\qquad\qquad\qquad\qquad-{\mathcal R^{m}_{nk}} + {\mathcal R^{k}_{n(m+k)}}.
			\end{eqnarray*}
		\end{small}
	\end{enumerate}
\end{proposition}
\section{$\mathcal{R}(p,q)-$ deformed Virasoro algebra}
In this section,   the $\mathcal{R}(p,q)$-deformed Virasoro algebra is realized as the central extension of the Witt algebra (\ref{RWa}). It is generated by operators acting as  follows 
{\begin{eqnarray*}
		L^{\mathcal{R}(p,q)}_n\varphi(z) &:=&\epsilon^{-z\partial_z}_1\,\big[z\partial_z-n\big]_{\mathcal{R}(p,q)}\,z^n\,\varphi(z)
\end{eqnarray*}
obeying the commutation relations}
\begin{eqnarray}\label{RV}
\Big[L^{\mathcal{R}(p,q)}_n,L^{\mathcal{R}(p,q)}_m\Big]_{\epsilon^{m-n}_1,\,\epsilon^{m-n}_2} = [m-n]_{\mathcal{R}(p,q)}\,L^{\mathcal{R}(p,q)}_{n+m} + \delta_{n+m,o}\,\mathcal{C}_{\mathcal{R}(p,q)}(p,q),
\end{eqnarray}

\begin{eqnarray}\label{RV1}
\Big[L^{\mathcal{R}(p,q)}_k,C_{\mathcal{R}(p,q)}(n)\Big]_{\epsilon^{-k}_1, \epsilon^{-k}_2}=0,
\end{eqnarray}
 wth the $\mathcal{R}(p,q)-$ deformed central term  $\mathcal{C}_{\mathcal{R}(p,q)}(p,q)$   given by

\begin{eqnarray}\label{Rct}
	C_{\mathcal{R}(p,q)}(n)=
	C(p,q)(\epsilon_1\,\epsilon_2)^{-2\,n}{[n]_{\mathcal{R}(p,q)}\over [2\,n]_{\mathcal{R}(p,q)}}\,[n-1]_{\mathcal{R}(p,q)}\,[n]_{\mathcal{R}(p,q)}\,[n+1]_{\mathcal{R}(p,q)},
	\end{eqnarray}
	where
	$C(p,q)$ is an arbitrary function of $(p,q).$
It is worth  noticing the realization of $\mathcal{R}(p,q)$-deformed Virasoro algebras
in terms of  concrete difference operators corresponding to  known deformed quantum algebras as follows:
	\begin{enumerate}
		\item[(i)] For $\epsilon_1=1$ and $\epsilon_2=q,$ we obtain the Virasoro algebra associated to the {\bf Arick} and {\bf Coon} deformation \cite{AC} with the generators $L^{q}_n$:
		{\begin{eqnarray*}
				L^{q}_n\varphi(z) &=&\big[z\partial_z-n\big]_{q}\,z^n\,\varphi(z)
			\end{eqnarray*}
			satisfying the commutation relations}
		\begin{eqnarray*}
			\,\Big[L^{q}_n, L^{q}_m\Big]_{1,q^{m-n}}&=&[m-n]_{q}\,L^{q}_{n+m} + \delta_{n+m,0}\,C_{q}(n),\\
			\,\Big[L^{q}_k, C_{q}(n)\Big]_{1,q^{-k}}&=&0,
		\end{eqnarray*}
		with \begin{eqnarray*}
			C_{q}(n)=
			C(q)q^{-2\,n}{[n]_{q}\over [2\,n]_{q}}\,[n-1]_{q}\,[n]_{q}\,[n+1]_{q}.
		\end{eqnarray*}
		\item[(ii)] The Virasoro algebra associated to the \textbf{Jagannathan-Srinivasa} deformation \cite{JS} is derived by taking $\epsilon_1=p$ and  $\epsilon_2=q,$ giving
		{\begin{eqnarray*}
			L^{p,q}_n\varphi(z) =\,p^{-z\partial_z}\,\big[z\partial_z-n\big]_{p,q}\,z^n\,\varphi(z),
		\end{eqnarray*}
		which induces the commutation relations:}
		\begin{eqnarray*}
			\,\Big[L^{p,q}_n, L^{p,q}_m\Big]_{p^{m-n},q^{m-n}}&=&[m-n]_{p,q}\,L^{p,q}_{n+m} + \delta_{n+m,0}\,C_{p,q}(n)\\►
			\,\Big[L^{p,q}_k, C_{p,q}(n)]_{p^{-k},\, q^{-k}}&=&0,
		\end{eqnarray*}
		with
		\begin{equation*}
			C_{p,q}(n)=
			C(p,q)(p\,q)^{-2\,n}{[n]_{p,q}\over [2\,n]_{p,q}}\,[n-1]_{p,q}\,[n]_{p,q}\,[n+1]_{p,q}.
		\end{equation*}
		\item[(ii)] Taking $\epsilon_1=p^{-1}$ and $\epsilon_2=q,$ we deduce the Virasoro algebra associated to the  \textbf{Chakrabarty- Jagannathan} deformation \cite{Chakrabarti&Jagan} with the generator action :
		{\begin{eqnarray*}
			L^{p^{-1},q}_n\varphi(z) =\,p^{z\partial_z}\,\big[z\partial_z-n\big]_{p^{-1},q}\,z^n\,\varphi(z),
		\end{eqnarray*}
		giving the commutation relations }
		\begin{eqnarray*}
			\,\Big[L^{p^{-1},q}_n, L^{p^{-1},q}_m\Big]_{p^{n-m},q^{m-n}}&=&[m-n]_{p^{-1},q}\,L^{p^{-1},q}_{n+m} + \delta_{n+m,0}\,C_{p^{-1},q}(n)\\
			\,\Big[L^{p^{-1},q}_k, C_{p^{-1},q}(n)]_{p^{k},\, q^{-k}}&=&0,
		\end{eqnarray*}
		with
		\begin{small}
			\begin{equation*}
				C_{p^{-1},q}(n)=
				C(p,q)(p^{-1}\,q)^{-2\,n}{[n]_{p^{-1},q}\over [2\,n]_{p^{-1},q}}\,[n-1]_{p^{-1},q}\,[n]_{p^{-1},q}\,[n+1]_{p^{-1},q}.
			\end{equation*}
		\end{small}
		\item[(iv)] The Virasoro algebra associated to the {\bf Hounkonnou-Ngompe} generalization of $q-$ Quesne deformation \cite{HN} can be obtained by putting $\epsilon_1=p$ and $\epsilon_2=q^{-1}$ yielding :  
		{\begin{eqnarray*}
			L^{p,q}_n\varphi(z) =\,p^{-z\partial_z-1}\,q\,\big[z\partial_z-n\big]_{p,q}\,z^n\,\varphi(z),
		\end{eqnarray*}
		with the commutation relations:}
		\begin{eqnarray*}
			\Big[L^{p,q}_n, \,L^{p,q}_m\Big]_{p^{m-n},q^{m-n}}&=&p\,q^{-1}[m-n]_{p,q}\,L^{p,q}_{n+m} + \delta_{n+m,0}\,C_{p,q}(n),\\
			\Big[L^{p,q^{-1}}_k,\, C_{p,q}(n)]_{p^{-k},\, q^{k}}&=&0,
		\end{eqnarray*}
		and
		\begin{equation*}
			C_{p,q}(n)=
			C(p,q)(p\,q^{-1})^{-2\,n-1}{[n]_{p,q}\over [2\,n]_{p,q}}\,[n-1]_{p,q}\,[n]_{p,q}\,[n+1]_{p,q}.
		\end{equation*}
	\end{enumerate}

In addition, let us mention that
the $q-$ deformed Virasoro algebra given in \cite{CILPP} is easily recovered by taking $\epsilon_1=q$ and $\epsilon_2=q^{-1}$ as follows:
{\begin{eqnarray*}
	L^{q}_n\varphi(z) =q\,\big[z\partial_z-n\big]_{q}\,z^n\,\varphi(z),
\end{eqnarray*}
with }
		\begin{eqnarray*}
			\,\Big[L^{q}_n, L^{q}_m\Big]_{q^{n-m},q^{m-n}}&=&[m-n]_{q}\,L^{q}_{n+m} + \delta_{n+m,0}\,C_{q}(n), \\
			\,\Big[L^{q}_k, C_{q}(n)\Big]_{q^{k},q^{-k}}&=&0
		\end{eqnarray*}
		and \begin{eqnarray*}
			C_{q}(n)=
			C(q){[n]_{q}\over [2\,n]_{q}}\,[n-1]_{q}\,[n]_{q}\,[n+1]_{q},
		\end{eqnarray*}
while the $(p,q)-$ deformed Virasoro algebra obtained in \cite{CJ} is deduced by setting $\epsilon_1=p$ and $\epsilon_2=q$ with the characteristics:
{\begin{eqnarray*}
	L^{p,q}_n\varphi(z) =\,p^{-z\partial_z}\,\big[z\partial_z-n\big]_{p,q}\,z^n\,\varphi(z),
\end{eqnarray*}}
 \begin{eqnarray*}
			\,\Big[L^{p,q}_n, L^{p,q}_m\Big]_{p^{n-m},q^{m-n}}&=&[m-n]_{p,q}\,L^{p,q}_{n+m} + \delta_{n+m,0}\,C_{p,q}(n),\\
			\,\Big[L^{p,q}_k, C_{p,q}(n)\Big]_{p^{k},q^{-k}}&=&0,
		\end{eqnarray*}
		and
		\begin{eqnarray*}
			C_{p,q}(n)=
			C(p,q)(p^{-1}\,q)^{-2\,n}{[n]_{p,q}\over [2\,n]_{p,q}}\,[n-1]_{p,q}\,[n]_{p,q}\,[n+1]_{p,q}.
		\end{eqnarray*}

\subsection{$\mathcal{R}(p,q)-$ deformed Korteweg-de Vries equation}
Let us now derive the $\mathcal{R}(p,q)-$ deformed Korteweg-de Vries (KdV) equation, and deduce its  particular  cases corresponding to known deformed quantum algebras. For that, we redefine the $\mathcal{R}(p,q)-$ deformed generators  as follows:
\begin{eqnarray*}
	\mathcal{L}^{\mathcal{R}(p,q)}_n\varphi(z):
	=\epsilon^{-z\partial_z}_2\,\big[z\partial_z-n\big]_{\mathcal{R}(p,q)}\,z^n\,\varphi(z),
\end{eqnarray*}
with the commutator  
\begin{eqnarray}\label{crv}
\Big[\mathcal{L}^{\mathcal{R}(p,q)}_n, \mathcal{L}^{\mathcal{R}(p,q)}_m\Big]=[m-n]_{\mathcal{R}(p,q)}\,\epsilon^{z\partial_z-m}_1\,\epsilon^{-z\partial_z+n}_2\,\mathcal{L}^{\mathcal{R}(p,q)}_{n+m} + \delta_{n+m,0}\,\mathcal{C}_{\mathcal{R}(p,q)}(n),
\end{eqnarray}
where
\begin{small}
	\begin{eqnarray*}
		\mathcal{C}_{\mathcal{R}(p,q)}(n)=\mathcal{C}(p,q)(\epsilon_1\,\epsilon_2)^{-2\,n}\,{\epsilon^{z\partial_z-n}_1\over\epsilon^{z\partial_z+n}_2}\,{[n]_{\mathcal{R}(p,q)}\over [2\,n]_{\mathcal{R}(p,q)}}\,[n-1]_{\mathcal{R}(p,q)}\,[n]_{\mathcal{R}(p,q)}\,[n+1]_{\mathcal{R}(p,q)}.
	\end{eqnarray*}
\end{small}
The $\mathcal{R}(p,q)-$ deformed current can thus  be expressed by:
\begin{equation*}
	w(\tau):=\displaystyle\sum_{n\in\mathbb{Z}
	}\,\mathcal{L}^{\mathcal{R}(p,q)}_n\,e^{i\,n\,\tau}.
\end{equation*}
We then arrive at the next statement.
\begin{theorem}
	The $\mathcal{R}(p,q)-$ deformed Korteweg-de Vries equation is written  as:
	\begin{eqnarray*}
		2\sin \tau\,\frac{dv}{dt}=\Theta\Big(e^{-2\tau\partial_x}\,v^2(x)-v(x)\,e^{2\tau\partial_x}\,u(x)\Big)-2\,\mathcal{C}_{\mathcal{R}(p,q)}\,\Theta^{3}\,\sinh 2\tau\partial_x\,v(x),
	\end{eqnarray*}
	where $\Theta=\big(\epsilon_1\epsilon_2\big)^{-1/2},$  $\Delta=\Big({\epsilon_2\over \epsilon_1}\Big)^{1/2}.$
\end{theorem}
{\it Proof.} From the commutation relation (\ref{crv}), 
\begin{eqnarray*}
	w(b)\,e^{-2\tau\partial_a}\,\delta(a-b)=e^{-2\tau\partial_a}\,w(a)\,\delta(a-b)
\end{eqnarray*}
and setting $\Theta=\big(\epsilon_1\epsilon_2\big)^{-1/2}$ and  $\Delta=\Big({\epsilon_2\over \epsilon_1}\Big)^{1/2},$ we obtain:
\begin{eqnarray*}
	\Big[w(a),w(b)\Big]&=& P\,\delta(a-b)\nonumber\\
	&=& \frac{2\pi\,i\Theta}{2\sin \tau}\Big( e^{-2\tau\partial_a}w(a)-w(a)e^{2\tau\partial_a}\Big)\Delta^{-2\,N}\delta(a-b)
	\nonumber\\
	&-&\Theta^{3}\mathcal{C}_{\mathcal{R}(p,q)}\frac{\sinh \tau\partial_a }{\sinh 2\tau\partial_a}\frac{\sinh \tau(\partial_a+i)\sinh \tau\partial_a \sinh \tau(\partial_a -i) }{\sin ^3\tau}\nonumber\\
	&\times& \Delta^{-2\,N}\delta(a-b),
\end{eqnarray*}
where $P$ stands for the  Hamiltonian operator.
Then, 
\begin{eqnarray}\label{RK1}
\frac{dw}{dt}&=&\frac{\Theta}{4\sin \tau}\Big( e^{-2\tau\partial_a}w(a)-w(a)e^{2\tau\partial_a}\Big)\Big(\Delta^{-2N}w(a)+w(a)\Delta^{-2N}\Big)\nonumber\\
&-& \frac{\Theta^{3}}{2}\mathcal{C}_{\mathcal{R}(p,q)}\frac{\sinh \tau\partial_a }{\sinh 2\tau\partial_a }\frac{\sinh \tau(\partial_a+i)\sinh \tau\partial_a \sinh \tau(\partial_a -i) }{\sin ^3\tau}\nonumber\\
&\times& \Big(\Delta^{-2N}w(a)+w(a)\Delta^{-2N}\Big).
\end{eqnarray} 
Setting  $w(x)=\Delta^{2N}\,v(x),$  we expand  (\ref{RK1}) as
\begin{eqnarray*}
	\frac{dw}{dt}&=&\frac{\Theta}{4\sin \tau}\Big( e^{-2\tau\partial_x}w(x)-w(x)e^{2\tau\partial_x}\Big)\Big(v(x)+w(x)\Delta^{-2N}\Big)\nonumber\\
	&-& \frac{\Theta^{3}}{2}\mathcal{C}_{\mathcal{R}(p,q)}\frac{\sinh \tau\partial_x }{\sinh 2\tau\partial_x }\frac{\sinh \tau(\partial_x+i)\sinh \tau\partial_x \sinh \tau(\partial_x -i) }{\sin ^3\tau}\nonumber\\
	&\times& \Big(v(x)+w(x)\Delta^{-2N}\Big)\nonumber\\
	&=&\frac{\Theta\Delta^{2N}}{4\sin \tau}\Big(e^{-2\tau\partial_x}\,v(x)-v(x)e^{2\tau\partial_x}\Big)\Big(v(x)+\Delta^{2N}v(x)\Delta^{-2N}\Big)\nonumber\\
	&-& \frac{\Theta^{3}}{2}\mathcal{C}_{\mathcal{R}(p,q)}\frac{\sinh \tau\partial_x }{\sinh 2\tau\partial_x }\frac{\sinh \tau(\partial_x+i)\sinh \tau\partial_x \sinh \tau(\partial_x -i) }{\sin ^3\tau}\nonumber\\
	&\times& \Big(v(x)+\Delta^{2N}\,v(x)\,\Delta^{-2N}\Big)\nonumber\\
	&=&\frac{\Theta\Delta^{2N}}{2\sin \tau}\Big(e^{-2\tau\partial_x}\,v(x)-v(x)e^{2\tau\partial_x}\Big)v(x)\nonumber\\
	&-& \Theta^{3}\mathcal{C}_{\mathcal{R}(p,q)}\frac{\sinh \tau\partial_x }{\sinh 2\tau\partial_x }\frac{\sinh \tau(\partial_x+i)\sinh \tau\partial_x \sinh \tau(\partial_x -i) }{\sin ^3\tau}\,v(x) 
\end{eqnarray*}
yielding the $\mathcal{R}(p,q)-$ deformed Korteweg-de Vries equation
\begin{eqnarray*}
	2\sin \tau\,\frac{dv}{dt}=\Theta\Big(e^{-2\tau\partial_x}\,v^2(x)-v(x)\,e^{2\tau\partial_x}\,v(x)\Big)-2\,\mathcal{C}_{\mathcal{R}(p,q)}\,\Theta^{3}\,\sinh 2\tau\partial_x\,v(x).
\end{eqnarray*} 
\cqfd

 For $\epsilon_1=q$ and $\epsilon_2=q^{-1},$ we recover the $q-$ deformed KdV equation \cite{CPPa}:
		\begin{eqnarray*}
			2\sin \tau\,\frac{dv}{dt}=e^{-2\tau\partial_x}\,v^2(x)-v(x)\,e^{2\tau\partial_x}\,v(x)-2\,\mathcal{C}_q\,\sinh 2\tau\partial_x\,v(x).
		\end{eqnarray*}
		The $(p,q)-$ deformed KdV equation given in \cite{CJ} can be obtained by taking $\epsilon_1=q$ and $\epsilon_2=p^{-1}:$ \begin{eqnarray*}
			\frac{dw}{dt}&=&\frac{\Theta}{4\sin \tau}\Big( e^{-2\tau\partial_a}w(a)-w(a)e^{2\tau\partial_a}\Big)\Big(\Delta^{-2N}w(a)+w(a)\Delta^{-2N}\Big)\nonumber\\
			&-& \frac{\Theta^{3}}{2}\,\mathcal{C}_{p,q}\,\frac{\sinh \tau\partial_a }{\sinh 2\tau\partial_a }\frac{\sinh \tau(\partial_a+i)\sinh \tau\partial_a \sinh \tau(\partial_a -i) }{\sin ^3\tau}\nonumber\\
			&\times& \Big(\Delta^{-2N}w(a)+w(a)\Delta^{-2N}\Big)
		\end{eqnarray*}
		with $\Theta=\Big(q\,p^{-1}\Big)^{1/2}$ and $\Delta=\Big(q\,p\Big)^{1/2}.$

 Now, we can  easily   deduce 
	relevant particular new Korteweg-de Vries (KdV) equations associated with  deformations   spread in the literature as follows:
	\begin{itemize}
		\item[(i)] For $\epsilon_1=1$ and $\epsilon_2=q,$ we deduct the KdV equation associated to the {\bf Arick} and {\bf Coon} deformation \cite{AC}:
		\begin{eqnarray*}
			2\sin \tau\,\frac{dv}{dt}=\Theta\Big(e^{-2\tau\partial_x}\,v^2(x)-v(x)\,e^{2\tau\partial_x}\Big)\,v(x)-2\,\mathcal{C}_q\,\Theta^3\,\sinh 2\tau\partial_x\,v(x)
		\end{eqnarray*}
		where $\Theta=q^{-1/2}.$
		\item[(ii)] The  Korteweg-de Vries equation associated to the \textbf{Jagannathan-Sirinivasa} deformation \cite{JS} can be established by putting $\epsilon_1=p$ and $\epsilon_2=q$: 
		\begin{eqnarray*}
			2\sin \tau\,\frac{dv}{dt}=\Theta\Big(e^{-2\tau\partial_x}\,v^2(x)-v(x)\,e^{2\tau\partial_x}\,v(x)\Big)-2\,\mathcal{C}_{p,q}\,\Theta^{3}\,\sinh 2\tau\partial_x\,v(x)
		\end{eqnarray*}
		with $\Theta=\Big(p\,q\Big)^{-1/2}.$
		\item[(iii)] Taking $\epsilon_1=p^{-1}$ and $\epsilon_2=q,$ we obtain the KdV equation associated to the \textbf{Chakrabarty-Jagannathan} deformation \cite{Chakrabarti&Jagan} : \begin{eqnarray*}
			2\sin \tau\,\frac{dv}{dt}=\Theta\Big(e^{-2\tau\partial_x}\,v^2(x)-v(x)\,e^{2\tau\partial_x}\,v(x)\Big)-2\,\mathcal{C}_{p^{-1},q}\,\Theta^{3}\,\sinh 2\tau\partial_x\,v(x)
		\end{eqnarray*}
		with $\Theta=\big(p^{-1}\,q\big)^{-1/2}.$
		\item [(iv)]The KdV equation corresponding to the \textbf{Hounkonnou-Ngompe} generalization of $q-$ Quesne \cite{HN} is deduced by taking $\epsilon_1=p$ and $\epsilon_2=q^{-1}:$ \begin{eqnarray*}
			2\sin \tau\,\frac{dv}{dt}=\Theta\Big(e^{-2\tau\partial_x}\,v^2(x)-v(x)\,e^{2\tau\partial_x}\,v(x)\Big)-2\,\mathcal{C}^Q_{p,q}\,\Theta^{3}\,\sinh 2\tau\partial_x\,v(x)
		\end{eqnarray*}
		where $\Theta=\big(p\,q^{-1}\big)^{-1/2}.$
	\end{itemize}
	\section{${\mathcal R}(p,q)-$ deformed Witt $n-$ algebra}
	 Let us consider  the operators defined by:
\begin{eqnarray}\label{to}
{\mathcal T}^{{\mathcal R}(p^{\alpha},q^{\alpha})}_m:=-{\mathcal D}_{{\mathcal R}(p^{\alpha},q^{\alpha})}\,z^{m+1},
\end{eqnarray}
where ${\mathcal D}_{{\mathcal R}(p^{\alpha},q^{\alpha})}$ is the ${\mathcal R}(p,q)-$ deformed derivative given by:
\begin{eqnarray*}
{\mathcal D}_{{\mathcal R}(p^{\alpha},q^{\alpha})}\big(\phi(z)\big)={p^{\alpha}-q^{\alpha}\over p^{\alpha\,P}-q^{\alpha\,Q}}{\mathcal R}(p^{\alpha\,P},q^{\alpha\,Q})\,{\phi(p^{\alpha}z)-\phi(q^{\alpha}z)\over p^{\alpha}-q^{\alpha}}.
\end{eqnarray*}
From (\ref{rpq number}), the operators (\ref{to}) take the form
\begin{eqnarray*}
{\mathcal T}^{{\mathcal R}(p^{\alpha},q^{\alpha})}_m=-[m+1]_{{\mathcal R}(p^{\alpha},q^{\alpha})}\,z^{m}.
\end{eqnarray*}
\begin{proposition}
	The operators (\ref{to}) satisfy the product relation \begin{eqnarray}\label{pre}
	&& {\mathcal T}^{{\mathcal R}(p^{\alpha},q^{\alpha})}_m.{\mathcal T}^{{\mathcal R}(p^{\beta},q^{\beta})}_n=-{\big(\epsilon^{\alpha+\beta}_1-\epsilon^{\alpha+\beta}_2\big)\epsilon^{-m\,\beta}_1\over \big(\epsilon^{\alpha}_1-\epsilon^{\alpha}_2\big)\big(\epsilon^{\beta}_1-\epsilon^{\beta}_2\big)}\,{\mathcal T}^{{\mathcal R}(p^{\alpha+\beta},q^{\alpha+\beta})}_{m+n} + {\epsilon^{(n+1)\beta}_2\over \epsilon^{\beta}_1-\epsilon^{\beta}_2}\, {\mathcal T}^{{\mathcal R}(p^{\alpha},q^{\alpha})}_{m+n}\cr&&\qquad\qquad\qquad\qquad\qquad\qquad\qquad\qquad\qquad\qquad\qquad\qquad + {\epsilon^{-m\,\beta}_1\,\epsilon^{(m+n+1)\alpha}_2\over \epsilon^{\alpha}_1-\epsilon^{\alpha}_2}\,{\mathcal T}^{{\mathcal R}(p^{\beta},q^{\beta})}_{m+n}
	\end{eqnarray}
	and the commutation relation
	\begin{eqnarray}\label{crto}
	&&\Big[{\mathcal T}^{{\mathcal R}(p^{\alpha},q^{\alpha})}_m, {\mathcal T}^{{\mathcal R}(p^{\beta},q^{\beta})}_n\Big]={\big(\epsilon^{\alpha+\beta}_1-\epsilon^{\alpha+\beta}_2\big)\big(\epsilon^{-n\,\alpha}_1-\epsilon^{-m\,\beta}_1\big)\over \big(\epsilon^{\alpha}_1-\epsilon^{\alpha}_2\big)\big(\epsilon^{\beta}_1-\epsilon^{\beta}_2\big)}\,{\mathcal T}^{{\mathcal R}(p^{\alpha+\beta},q^{\alpha+\beta})}_{m+n}\cr&&\qquad -{\epsilon^{(m+n+1)\beta}_2\big(\epsilon^{-n\,\alpha}_1-\epsilon^{-m\,\beta}_2\big)\over \epsilon^{\beta}_1-\epsilon^{\beta}_2}\, {\mathcal T}^{{\mathcal R}(p^{\alpha},q^{\alpha})}_{m+n} + {\epsilon^{(m+n+1)\alpha}_2\big(-\epsilon^{-m\,\beta}_1-\epsilon^{-n\,\alpha}_2\big)\over \epsilon^{\alpha}_1-\epsilon^{\alpha}_2}\,{\mathcal T}^{{\mathcal R}(p^{\beta},q^{\beta})}_{m+n}.
	\end{eqnarray}
\end{proposition}

Taking $\alpha=\beta=1,$ we obtain:
\begin{small}
\begin{eqnarray*}\label{crto1}
&&\Big[{\mathcal T}^{{\mathcal R}(p,q)}_m, {\mathcal T}^{{\mathcal R}(p,q)}_n\Big]={\big(\epsilon^{-n}_1-\epsilon^{-m}_1\big)\over \big(\epsilon_1-\epsilon_2\big)}\,[2]_{{\mathcal R}(p,q)}{\mathcal T}^{{\mathcal R}(p^{2},q^{2})}_{m+n}\cr&&\qquad\qquad\qquad\qquad\qquad\qquad\qquad\qquad\qquad-{\epsilon^{m+n+1}_2\over \epsilon_1-\epsilon_2}\Big(\big(\epsilon^{-n}_1-\epsilon^{-m}_2\big)-\big(\epsilon^{-m}_1-\epsilon^{-n}_2\big)\Big) {\mathcal T}^{{\mathcal R}(p,q)}_{m+n}.
\end{eqnarray*}
\end{small}
Note that for ${\mathcal R}(q,1)=1,$ involving $\epsilon_1=q$ and $\epsilon_2=q,$ we obtain the result given in \cite{NZ}:
\begin{eqnarray*}
\big[{\mathcal T}^{q}_m, {\mathcal T}^{q}_n\big]&=&\big([m+1]_{q}- [n+1]_{q}\big){\mathcal T}^{q}_{m+n}\nonumber\\
&=& \big([-n]_q-[-m]_q\big)\big([2]_q{\mathcal T}^{q^2}_{m+n}-{\mathcal T}^{q}_{m+n}\big).
\end{eqnarray*}
We consider the $n-$ bracket defined by:
\begin{eqnarray*}
\Big[{\mathcal T}^{{\mathcal R}(p^{\alpha_1},q^{\alpha_1})}_{m_1},\cdots,{\mathcal T}^{{\mathcal R}(p^{\alpha_n},q^{\alpha_n})}_{m_n}
\Big]:=\Gamma^{i_1 \cdots i_n}_{1 \cdots n}\,{\mathcal T}^{{\mathcal R}(p^{\alpha_{i_1}},q^{\alpha_{i_1}})}_{m_{i_1}} \cdots {\mathcal T}^{{\mathcal R}(p^{\alpha_{i_n}},q^{\alpha_{i_n}})}_{m_{i_n}}
\end{eqnarray*}
where $\Gamma^{i_1 \cdots i_n}_{1 \cdots n}$ is the L\'evi-Civit\'a symbol given by:
\begin{eqnarray*}
\Gamma^{j_1 \cdots j_p}_{i_1 \cdots i_p}= det\left( \begin{array} {ccc}
\delta^{j_1}_{i_1} &\cdots&  \delta^{j_1}_{i_p}   \\ 
\vdots && \vdots \\
\delta^{j_p}_{i_1} & \cdots& \delta^{j_p}_{i_p}
\end {array} \right) .
\end{eqnarray*}
Focussing on the case with the same ${\mathcal R}(p^{\alpha},q^{\alpha})$ leads to
\begin{eqnarray*}
\Big[{\mathcal T}^{{\mathcal R}(p^{\alpha},q^{\alpha})}_{m_1},\cdots,{\mathcal T}^{{\mathcal R}(p^{\alpha},q^{\alpha})}_{m_n}
\Big]=\Gamma^{1\cdots n}_{1\cdots n}\,{\mathcal T}^{{\mathcal R}(p^{\alpha},q^{\alpha})}_{m_{1}}\cdots {\mathcal T}^{{\mathcal R}(p^{\alpha},q^{\alpha})}_{m_{n}}.
\end{eqnarray*}
Taking $\alpha=\beta$ in the relation (\ref{crto}), we obtain:
\begin{eqnarray*}\label{crtob}
	&&\Big[{\mathcal T}^{{\mathcal R}(p^{\alpha},q^{\alpha})}_m, {\mathcal T}^{{\mathcal R}(p^{\alpha},q^{\alpha})}_n\Big]={\big(\epsilon^{-n\alpha}_1-\epsilon^{-m\alpha}_1\big)\over \big(\epsilon^{\alpha}_1-\epsilon^{\alpha}_2\big)}\,[2]_{{\mathcal R}(p^{\alpha},q^{\alpha})}{\mathcal T}^{{\mathcal R}(p^{2\alpha},q^{2\alpha})}_{m+n}\cr&&\qquad\qquad\qquad\qquad\qquad\qquad-{\epsilon^{(m+n+1)\alpha}_2\over \epsilon^{\alpha}_1-\epsilon^{\alpha}_2}\Big(\big(\epsilon^{-n\alpha}_1-\epsilon^{-m\alpha}_1\big)+\big(\epsilon^{-n\alpha}_2-\epsilon^{-m\alpha}_2\big)\Big) {\mathcal T}^{{\mathcal R}(p^{\alpha},q^{\alpha})}_{m+n}.
\end{eqnarray*}
The $n-$ bracket can then be rewritten as follows:
 \begin{eqnarray}\label{crna}
 &&\Big[{\mathcal T}^{{\mathcal R}(p^{\alpha},q^{\alpha})}_{m_1},\cdots, {\mathcal T}^{{\mathcal R}(p^{\alpha},q^{\alpha})}_{m_n}\Big]={(-1)^{n+1}\over \big(\epsilon^{\alpha}_1-\epsilon^{\alpha}_2\big)^{n-1}}\Big( H^n_{\alpha}\,[n]_{{\mathcal R}(p^{\alpha},q^{\alpha})}{\mathcal T}^{{\mathcal R}(p^{n\alpha},q^{n\alpha})}_{m_1+\cdots+m_n} - [n-1]_{{\mathcal R}(p^{\alpha},q^{\alpha})}\cr&&\qquad\qquad\qquad\qquad\qquad\qquad\qquad\qquad \times \epsilon^{\alpha\big(\sum_{l=1}^{n}m_l+1\big)}_2\big(H^n_{\alpha}+ M^n_{\alpha}\big){\mathcal T}^{{\mathcal R}(p^{(n-1)\alpha},q^{(n-1)\alpha})}_{m_1+\cdots+m_n}\Big),
 \end{eqnarray}
 where 
 \begin{small}
 \begin{eqnarray*}
 H^n_{\alpha}&=& \epsilon^{-\alpha(n-1)\sum_{s=1}^{n}m_s}_1\Big(\big(\epsilon^{\alpha}_1-\epsilon^{\alpha}_2\big)^{n\choose 2}\prod_{1\leq j < k \leq n}\Big([m_k]_{{\mathcal R}(p^{\alpha},q^{\alpha})}-[m_j]_{{\mathcal R}(p^{\alpha},q^{\alpha})}\Big)\cr&+&\prod_{1\leq j < k \leq n}\Big(\epsilon^{\alpha\,m_k}_2-\epsilon^{\alpha\,m_j}_2\Big)\Big)
 \end{eqnarray*}
 and 
 \begin{eqnarray*}
 &&M^{n}_{\alpha}
  =\epsilon^{-\alpha(n-1)\sum_{s=1}^{n}m_s}_2\Big(\big(\epsilon^{\alpha}_1-\epsilon^{\alpha}_2\big)^{n\choose 2}\prod_{1\leq j < k \leq n}\Big([m_k]_{{\mathcal R}(p^{\alpha},q^{\alpha})}-[m_j]_{{\mathcal R}(p^{\alpha},q^{\alpha})}\Big)\cr&&\qquad\qquad\qquad\qquad\qquad\qquad\qquad\qquad\qquad\qquad\qquad\qquad\qquad+(-1)^{n-1}\prod_{1\leq j < k \leq n}\Big(\epsilon^{\alpha\,m_k}_1-\epsilon^{\alpha\,m_j}_1\Big)\Big).
 \end{eqnarray*}
\end{small}
Note that the  $q-$ Witt $n-$ algebra obtained in \cite{WYWZ} corresponds to the particular case of ${\mathcal R}(q,1)=1:$
\begin{eqnarray*}
\Big[{\mathcal T}^{q^{\alpha}}_{m_1},\cdots, {\mathcal T}^{q^{\alpha}}_{m_n}\Big]={(-1)^{n+1}H^n_{\alpha}\over (q^{\alpha}-1)^{n-1}}\Big([n]_{q^{\alpha}}{\mathcal T}^{q^{n\alpha}}_{m_1+\cdots+m_n} - [n-1]_{q^{\alpha}}\,{\mathcal T}^{q^{(n-1)\alpha}}_{m_1+\cdots+m_n}\Big),
\end{eqnarray*}
with 
$$H^{n}_{\alpha}
=(q^{\alpha}-1)^{n\choose 2}\,q^{-\alpha(n-1)\sum_{s=1}^{n}m_s}\,\prod_{1\leq j < k \leq n}\Big([m_k]_{q^{\alpha}}-[m_j]_{q^{\alpha}}\Big).$$
Taking $n=3$ in the relation (\ref{crna}), we obtain the ${\mathcal R}(p,q)-$ Witt $3-$ algebra:
\begin{eqnarray*}
&&\Big[{\mathcal T}^{{\mathcal R}(p^{\alpha},q^{\alpha})}_{m_1},{\mathcal T}^{{\mathcal R}(p^{\alpha},q^{\alpha})}_{m_2}, {\mathcal T}^{{\mathcal R}(p^{\alpha},q^{\alpha})}_{m_3}\Big]={1\over \big(\epsilon^{\alpha}_1-\epsilon^{\alpha}_2\big)^{2}}\Big( H^3_{\alpha}\,[3]_{{\mathcal R}(p^{\alpha},q^{\alpha})}{\mathcal T}^{{\mathcal R}(p^{3\alpha},q^{3\alpha})}_{m_1+\cdots+m_3} - [n-1]_{{\mathcal R}(p^{\alpha},q^{\alpha})}\cr&&\qquad\qquad\qquad\qquad\qquad\qquad\qquad\qquad\qquad\quad \times \epsilon^{\alpha\big(\sum_{l=1}^{3}m_l+1\big)}_2\big(H^3_{\alpha}+ M^3_{\alpha}\big){\mathcal T}^{{\mathcal R}(p^{2\alpha},q^{2\alpha})}_{m_1+\cdots+m_3}\Big),
\end{eqnarray*}
where 
\begin{eqnarray*}
	&&H^3_{\alpha}= \big(\epsilon^{\alpha}_1-\epsilon^{\alpha}_2\big)^{3}\,\epsilon^{-2\alpha(m_1+m_2+m_3)}_1\,\Big([m_2]_{{\mathcal R}(p^{\alpha},q^{\alpha})}-[m_1]_{{\mathcal R}(p^{\alpha},q^{\alpha})}\Big)\cr&&\qquad\qquad\qquad\qquad\qquad\times \Big([m_3]_{{\mathcal R}(p^{\alpha},q^{\alpha})}-[m_1]_{{\mathcal R}(p^{\alpha},q^{\alpha})}\Big)\Big([m_3]_{{\mathcal R}(p^{\alpha},q^{\alpha})}-[m_2]_{{\mathcal R}(p^{\alpha},q^{\alpha})}\Big)\cr&&\qquad\qquad\qquad\qquad+\epsilon^{-2\alpha(m_1+m_2+m_3)}_1\Big(\epsilon^{\alpha\,m_2}_2-\epsilon^{\alpha\,m_1}_2\Big)\Big(\epsilon^{\alpha\,m_3}_2-\epsilon^{\alpha\,m_1}_2\Big)\Big(\epsilon^{\alpha\,m_3}_2-\epsilon^{\alpha\,m_2}_2\Big)
\end{eqnarray*}
and 
\begin{eqnarray*}
	&&M^{3}_{\alpha}=
	\big(\epsilon^{\alpha}_1-\epsilon^{\alpha}_2\big)^{3}\,\epsilon^{-2\alpha(m_1+m_2+m_3)}_2\,\Big([m_2]_{{\mathcal R}(p^{\alpha},q^{\alpha})}-[m_1]_{{\mathcal R}(p^{\alpha},q^{\alpha})}\Big)\cr&&\qquad\qquad\qquad\qquad\quad\qquad\times\Big([m_3]_{{\mathcal R}(p^{\alpha},q^{\alpha})}-[m_1]_{{\mathcal R}(p^{\alpha},q^{\alpha})}\Big)\Big([m_3]_{{\mathcal R}(p^{\alpha},q^{\alpha})}-[m_2]_{{\mathcal R}(p^{\alpha},q^{\alpha})}\Big)\cr&&\qquad\qquad\qquad\qquad+\epsilon^{-2\alpha(m_1+m_2+m_3)}_2\Big(\epsilon^{\alpha\,m_2}_1-\epsilon^{\alpha\,m_1}_1\Big)\Big(\epsilon^{\alpha\,m_3}_1-\epsilon^{\alpha\,m_1}_1\Big)\Big(\epsilon^{\alpha\,m_3}_1-\epsilon^{\alpha\,m_2}_1\Big).
\end{eqnarray*}
\subsection{A toy model for ${\mathcal R}(p,q)-$ Virasoro constraints }
	In this section, we study a toy model for the ${\mathcal R}(p,q)-$ deformed Virasoro constraints, which  play an important role in the study of matrix models. Let us consider the generating function with infinitely many	parameters
 given as follows \cite{NZ}: $$Z^{toy}(t)=\int \, \,x^{\gamma}\,\exp\left(\displaystyle\sum_{s=0}^{\infty}{t_s\over s!}x^s\right)\,dx,$$
which encodes many different integrals.
The following property holds for the ${\mathcal R}(p,q)-$ deformed derivative
\begin{eqnarray*}
\int_{{\mathbb R}}{\mathcal D}_{{\mathcal R}(p^{\alpha},q^{\alpha})}f(x)d\,x={h(p^{\alpha},q^{\alpha})\over \epsilon^{\alpha}_1-\epsilon^{\alpha}_2}\Big(\int_{-\infty}^{+\infty}{f(\epsilon^{\alpha}_1\,x)\over x}dx -\int_{-\infty}^{+\infty}{f(\epsilon^{\alpha}_2\,x)\over x}dx\Big)=0,
\end{eqnarray*} 
where 
$$h(p^{\alpha},q^{\alpha})={p^{\alpha}-q^{\alpha}\over p^{P^{\alpha}}-q^{Q^{\alpha}}}{\mathcal R}\big(p^{P^{\alpha}},q^{Q^{\alpha}}\big).$$
For $f(x)=x^{m+\gamma+1}\,\exp\left(\displaystyle\sum_{s=0}^{\infty}{t_s\over s!}x^s\right),$ we have
\begin{eqnarray*}
\int_{-\infty}^{+\infty}{\mathcal D}_{{\mathcal R}(p^{\alpha},q^{\alpha})}\left(x^{m+\gamma+1}\,\exp\left(\sum_{s=0}^{\infty}{t_s\over s!}x^s\right)\right)d\,x=0.
\end{eqnarray*}
Considering the following expansion:
\begin{eqnarray*}
\exp\left(\displaystyle\sum_{s=0}^{\infty}{t_s\over s!}x^s\right)=\sum_{n=0}^{\infty}B_n(t_1,\cdots,t_n){x^n\over n!},
\end{eqnarray*}
where $B_n$ is the Bell polynomials, we get:
\begin{eqnarray*}
&&{\mathcal D}_{{\mathcal R}(p^{\alpha},q^{\alpha})}\left(x^{m+\gamma+1}\,\exp\left(\displaystyle\sum_{s=0}^{\infty}{t_s\over s!}x^s\right)\right)=\Big([m+1+\gamma]_{{\mathcal R}(p^{\alpha},q^{\alpha})}\,{x^m\over \epsilon^{\alpha\,m}_1}\cr&&\qquad\qquad\qquad\qquad\qquad + h(p^{\alpha},q^{\alpha}){\epsilon^{\alpha(m+1+\gamma)}_2\over \epsilon^{\alpha}_1 - \epsilon^{\alpha}_2}\sum_{k=1}^{\infty}{B_k(t^{\alpha}_1,\cdots,t^{\alpha}_k)\over k!}x^{k+m}\Big)x^{\gamma}\exp\left(\displaystyle\sum_{s=0}^{\infty}{t_s\over s!}x^s\right),
\end{eqnarray*}
where $t^{\alpha}_k=(\epsilon^{\alpha\,k}_1-\epsilon^{\alpha\,k}_2)t_k.$ Then,  using the constraints on the partition function,
$${\mathbb T}^{{\mathcal R}(p^{\alpha},q^{\alpha})}_m\,Z^{(toy)}(t)=0,\quad m\geq 0,$$
 we obtain 
 \begin{eqnarray*}\label{tom}
 &&{\mathbb T}^{{\mathcal R}(p^{\alpha},q^{\alpha})}_m=[m+1+\gamma]_{{\mathcal R}(p^{\alpha},q^{\alpha})}\,m!\, \epsilon^{-\alpha\,m}_1\,{\partial\over \partial t_m}\cr&&\qquad\qquad\qquad\qquad\qquad + h(p^{\alpha},q^{\alpha}){\epsilon^{\alpha(m+1+\gamma)}_2\over \epsilon^{\alpha}_1 - \epsilon^{\alpha}_2}\sum_{k=1}^{\infty}{(k+m)!\over k!}B_k(t^{\alpha}_1,\cdots,t^{\alpha}_k){\partial\over \partial t_{k+m}}.
 \end{eqnarray*}
   Setting $\bar{m}=m+1+\gamma,\quad \bar{n}=n+1+\gamma,$ and using the substitution, $$n!\,{\partial\over \partial t_n}\longleftrightarrow x^n,$$
we get
\begin{small}
\begin{eqnarray*}
{\mathbb T}^{{\mathcal R}(p^{\alpha},q^{\alpha})}_m.{\mathbb T}^{{\mathcal R}(p^{\beta},q^{\beta})}_n&=&[\bar{n}]_{{\mathcal R}(p^{\beta},q^{\beta})}\epsilon^{-\alpha\,m-\beta\,n}_1\,{\mathbb T}^{{\mathcal R}(p^{\alpha},q^{\alpha})}_{n+m} +{h(p^{\alpha},q^{\alpha})h(p^{\beta},q^{\beta})\epsilon^{\alpha\bar{m}+\beta\bar{n}}_2\over (\epsilon^{\alpha}_1-\epsilon^{\alpha}_2)(\epsilon^{\beta}_1-\epsilon^{\alpha}_2)} \nonumber\\ &\times& \sum_{k=1}^{\infty}\sum_{l=1}^{\infty}B_k(t^{\alpha}_1,\cdots,t^{\alpha}_k)B_l(t^{\beta}_1,\cdots,t^{\beta}_l) {1\over k!}{1\over l!}x^{k+l+m+n}\nonumber\\ &+& {[\bar{m}]_{{\mathcal R}(p^{\alpha},q^{\alpha})}\epsilon^{-\alpha\,m}_1\epsilon^{\beta\bar{m}}_2h(p^{\beta},q^{\beta})\over \epsilon^{\beta}_1-\epsilon^{\beta}_2}\sum_{k=1}^{\infty}B_k(t^{\alpha}_1,\cdots,t^{\alpha}_k){1\over k!}x^{m+k+n}\nonumber\\
&+& {[\bar{n}]_{{\mathcal R}(p^{\beta},q^{\beta})}\epsilon^{-\alpha\,n}_1\epsilon^{\beta\bar{n}}_2h(p^{\alpha},q^{\alpha})\over \epsilon^{\alpha}_1-\epsilon^{\alpha}_2}\sum_{k=1}^{\infty}B_k(t^{\beta}_1,\cdots,t^{\beta}_k){1\over k!}x^{n+k+m}\nonumber\\
&-&{[\bar{n}]_{{\mathcal R}(p^{\beta},q^{\beta})}h(p^{\alpha},q^{\alpha})\epsilon^{\alpha\bar{m}+\beta\bar{n}}_2\over \epsilon^{\alpha\,m+\beta\,n}_1\big(\epsilon^{\alpha}_1-\epsilon^{\alpha}_2\big)}\sum_{k=1}^{\infty}{1\over k!}B_k(t_1,\cdots,t_k)x^{k+m+n}.
\end{eqnarray*}
\end{small}
After computation, we get \begin{eqnarray*}\label{crtom}
&&{\mathbb T}^{{\mathcal R}(p^{\alpha},q^{\alpha})}_m.{\mathbb T}^{{\mathcal R}(p^{\beta},q^{\beta})}_n\sim \,{\big(\epsilon^{\alpha+\beta}_1-\epsilon^{\alpha+\beta}_2\big)\epsilon^{-m\,\beta}_1\over \epsilon^{\alpha\,m+\beta\,n}_1\big(\epsilon^{\alpha}_1-\epsilon^{\alpha}_2\big)\big(\epsilon^{\beta}_1-\epsilon^{\beta}_2\big)}\,{\mathbb T}^{{\mathcal R}(p^{\alpha+\beta},q^{\alpha+\beta})}_{m+n} - {\epsilon^{(n+1)\beta}_2\over \epsilon^{\alpha\,m+\beta\,n}_1\big(\epsilon^{\beta}_1-\epsilon^{\beta}_2\big)}\, {\mathbb T}^{{\mathcal R}(p^{\alpha},q^{\alpha})}_{m+n}\cr&&\qquad\qquad\qquad\qquad\qquad\qquad\qquad\qquad\qquad\qquad\qquad\qquad - {\epsilon^{-m\,\beta}_1\,\epsilon^{(m+n+1)\alpha}_2\over \epsilon^{\alpha\,m+\beta\,n}_1\big(\epsilon^{\alpha}_1-\epsilon^{\alpha}_2\big)}\,{\mathcal T}^{{\mathcal R}(p^{\beta},q^{\beta})}_{m+n},
\end{eqnarray*}
where $\sim$ denotes the equivalence.
\subsection{relevant particular cases}
\subsubsection{Witt $n-$ algebra associated to the Arick and Coon deformation\cite{AC}}
	Putting ${\mathcal R}(q,1)=1,$ we obtain the $q-$ deformed Witt $n-$ algebra associated to the {\bf Arick} and {\bf Coon } deformation \cite{AC}. We define the $q-$ deformed operators as follows:
\begin{eqnarray}\label{qot}
{\mathcal T}^{q^{\alpha}}_m:=-{\mathcal D}_{q^{\alpha}}\,z^{m+1},
\end{eqnarray}
where ${\mathcal D}_{q^{\alpha}}$ is the $q-$ deformed derivative given by:
\begin{eqnarray*}
{\mathcal D}_{q^{\alpha}}\big(\phi(z)\big)={\phi(q\,z)-\phi(q^{-1}\,z)\over q-q^{-1}}
\end{eqnarray*}
and the $q-$ deformed number is
\begin{eqnarray*}
[n]_{q^{\alpha}}={q^{\alpha\,n}-q^{-\alpha\,n}\over q^{\alpha}-q^{-\alpha}}.
\end{eqnarray*}
Thus, the operators (\ref{qot}) are reduced to
\begin{eqnarray*}
{\mathcal T}^{q^{\alpha}}_m=-[m+1]_{q^{\alpha}}\,z^m
\end{eqnarray*}
and  satisfy  the relation
\begin{eqnarray*}
\Big[{\mathcal T}^{q^{\alpha}}_m,\, {\mathcal T}^{q^{\beta}}_n\Big]=[m+1]_{q^{\alpha}}\,{\mathcal T}^{q^{\beta}}_{m+n}- [n+1]_{q^{\beta}}{\mathcal T}^{q^{\alpha}}_{m+n}.
\end{eqnarray*}
Using the $q-$ number, we obtain  
\begin{eqnarray*}
{\mathcal T}^{q^{\alpha}}_m. {\mathcal T}^{q^{\beta}}_n=-{(q^{\alpha+\beta}-q^{-\alpha-\beta})q^{-\beta\,m}\over (q^{\alpha}-q^{-\alpha})(q^{\beta}-q^{-\beta})}T^{q^{\alpha+\beta}}_{m+n}+{q^{-\beta(n+1)}\over q^{\beta}-q^{-\beta}}{\mathcal T}^{q^{\alpha}}_{m+n}+{q^{-\alpha(m+n+1)-\beta\,m}\over q^{\alpha}-q^{-\alpha}}{\mathcal T}^{q^{\beta}}_{m+n}
\end{eqnarray*}
and the commutation relation
\begin{small}
\begin{eqnarray}\label{qcrto}
&&\big[{\mathcal T}^{q^{\alpha}}_m, {\mathcal T}^{q^{\beta}}_n\big]={(q^{\alpha+\beta}-q^{-\alpha-\beta})\big(q^{-\alpha\,n}-q^{-\beta\,m}\big)\over (q^{\alpha}-q^{-\alpha})(q^{\beta}-q^{-\beta})}{\mathcal T}^{q^{\alpha+\beta}}_{m+n} + {q^{-\beta(n+1)}-q^{-\beta(m+n+1)-\alpha\,n}\over q^{\beta}-q^{-\beta}}{\mathcal T}^{q^{\alpha}}_{m+n}\cr&&\qquad\qquad\qquad\qquad\qquad\qquad\qquad\qquad\qquad\qquad\qquad + {q^{-\alpha(m+n+1)-\beta\,m}-q^{-\alpha(m+1)}\over q^{\alpha}-q^{-\alpha}}{\mathcal T}^{q^{\beta}}_{m+n}.
\end{eqnarray}
\end{small}
Taking $\alpha=\beta=1$ in (\ref{qcrto}), we obtain
\begin{eqnarray*}
\big[{\mathcal T}^{q}_m, {\mathcal T}^{q}_n\big]={\big(q^{-n}-q^{-m}\big)\over q-q^{-1}}\Big([2]_{q}{\mathcal T}^{q^{2}}_{m+n} + q^{-1}\big(1-q^{-(m+n)}\big){\mathcal T}^{q}_{m+n}\Big),
\end{eqnarray*}
which can be rewritten as:
\begin{eqnarray}\label{al}
\big[{\mathcal T}^{q}_m, {\mathcal T}^{q}_n\big]={q^m\over q+1}\{m-n\}_{q}\Big(\{2\}_{q^2}{\mathcal T}^{q^{2}}_{m+n}-q^{-1}(q-1)\{-m-n\}_{q}{\mathcal T}^{q}_{m+n}\Big),
\end{eqnarray}
where
\begin{eqnarray*}
\{x\}={q^x-1\over q-1}.
\end{eqnarray*}
Taking the limit $q\longrightarrow 1,$ the algebra (\ref{al}) gives the Witt algebra.

	The $n-$ bracket with the same $q^{\alpha}$ is given by:
\begin{eqnarray*}
\Big[{\mathcal T}^{q^{\alpha}}_{m_1},\cdots,{\mathcal T}^{q^{\alpha}}_{m_n}
\Big]=\Gamma^{1\cdots n}_{1\cdots n}\,{\mathcal T}^{q^{\alpha}}_{m_{1}} \cdots {\mathcal T}^{q^{\alpha}}_{m_{n}}.
\end{eqnarray*}
For $\beta=\alpha,$ the relation (\ref{qcrto}) takes the following form:
\begin{small}
\begin{eqnarray*}
&&\big[{\mathcal T}^{q^{\alpha}}_{m_1}, {\mathcal T}^{q^{\alpha}}_{m_2}\big]={-\big(q^{-\alpha\,m_1}-q^{-\alpha\,m_2}\big)\over q^{\alpha}-q^{-\alpha}}\Big([2]_{q^{\alpha}}{\mathcal T}^{q^{2\alpha}}_{m_1+m_2} - q^{-\alpha}\big(1-q^{-\alpha(m_1+m_2)}\big){\mathcal T}^{q^{\alpha}}_{m_1+m_2}\Big).
\end{eqnarray*}
\end{small}
By induction, we deduce the $n-$ bracket as:
\begin{eqnarray}\label{crnaAC}
&&\Big[{\mathcal T}^{q^{\alpha}}_{m_1},\cdots, {\mathcal T}^{q^{\alpha}}_{m_n}\Big]={(-1)^{n+1}\over (q^{\alpha}-q^{-\alpha})^{n-1}}\Big( H^n_{\alpha}\,[n]_{q^{\alpha}}{\mathcal T}^{q^{n\alpha}}_{m_1+\cdots+m_n} - [n-1]_{q^{\alpha}}\cr&&\qquad\qquad\qquad\qquad\qquad\qquad\qquad\qquad\qquad \times q^{-\alpha\big(\sum_{l=1}^{n}m_l+1\big)}\big(H^n_{\alpha}+ M^n_{\alpha}\big){\mathcal T}^{q^{(n-1)\alpha}}_{m_1+\cdots+m_n}\Big),
\end{eqnarray}
where 
\begin{small}
	\begin{eqnarray*}
H^n_{\alpha}=q^{-\alpha(n-1)\sum_{s=1}^{n}m_s}\,\Big( (q^{\alpha}-q^{-\alpha})^{n\choose 2}\prod_{1\leq j < k \leq n}\Big([m_k]_{q^{\alpha}}-[m_j]_{q^{\alpha}}\Big)
+\prod_{1\leq j < k \leq n}\Big(q^{-\alpha\,m_k}-q^{-\alpha\,m_j}\Big)\Big)
	\end{eqnarray*}
	and 
	\begin{eqnarray*}
		&&M^{n}_{\alpha}
		=q^{\alpha(n-1)\sum_{s=1}^{n}m_s}\Big((q^{\alpha}-q^{-\alpha}\big)^{n\choose 2}\,\prod_{1\leq j < k \leq n}\Big([m_k]_{q^{\alpha}}-[m_j]_{q^{\alpha}}\Big)\cr&&\qquad\qquad\qquad\qquad\qquad\qquad\qquad\qquad\qquad\qquad\qquad+(-1)^{n-1}\prod_{1\leq j < k \leq n}\Big(q^{\alpha\,m_k}-q^{\alpha\,m_j}\Big)\Big).
	\end{eqnarray*}
\end{small}
Taking $n=3$ in (\ref{crnaAC}), we obtain the  Witt $3-$ algebra related to the {\bf Arick} and {\bf Coon} deformation \cite{AC}:
\begin{eqnarray*}
	&&\Big[{\mathcal T}^{q^{\alpha}}_{m_1},{\mathcal T}^{q^{\alpha}}_{m_2}, {\mathcal T}^{q^{\alpha}}_{m_3}\Big]={1\over \big(q^{\alpha}-q^{-\alpha}\big)^{2}}\Big( H^3_{\alpha}[3]_{q^{\alpha}}{\mathcal T}^{q^{3\alpha}}_{m_1+\cdots+m_3} - [n-1]_{q^{\alpha}}\cr&&\qquad\qquad\qquad\qquad\qquad\qquad\qquad\qquad\qquad\quad \times q^{-\alpha\big(\sum_{l=1}^{3}m_l+1\big)}\big(H^3_{\alpha}+ M^3_{\alpha}\big){\mathcal T}^{q^{2\alpha}}_{m_1+\cdots+m_3}\Big),
\end{eqnarray*}
where 
\begin{small}
\begin{eqnarray*}
	&&H^3_{\alpha}= \big(q^{\alpha}-q^{-\alpha}\big)^{3}\,q^{-2\alpha(m_1+m_2+m_3)}\,\Big([m_2]_{q^{\alpha}}-[m_1]_{q^{\alpha}}\Big)\Big([m_3]_{q^{\alpha}}-[m_1]_{q^{\alpha}}\Big)\Big([m_3]_{q^{\alpha}}-[m_2]_{q^{\alpha}}\Big)\cr&&\qquad\qquad\qquad+q^{-2\alpha(m_1+m_2+m_3)}\Big(q^{-\alpha\,m_2}-q^{-\alpha\,m_1}\Big)\Big(q^{-\alpha\,m_3}-q^{-\alpha\,m_1}\Big)\Big(q^{-\alpha\,m_3}-q^{-\alpha\,m_2}\Big)
\end{eqnarray*}
and 
\begin{eqnarray*}
	&&M^{3}_{\alpha}=
	\big(q^{\alpha}-q^{-\alpha}\big)^{3}\,q^{2\alpha(m_1+m_2+m_3)}\,\Big([m_2]_{q^{\alpha}}-[m_1]_{q^{\alpha}}\Big)\Big([m_3]_{q^{\alpha}}-[m_1]_{q^{\alpha}}\Big)\Big([m_3]_{q^{\alpha}}-[m_2]_{q^{\alpha}}\Big)\cr&&\qquad\qquad\qquad\qquad\qquad+q^{2\alpha(m_1+m_2+m_3)}\Big(q^{\alpha\,m_2}-q^{\alpha\,m_1}\Big)\Big(q^{\alpha\,m_3}-q^{\alpha\,m_1}\Big)\Big(q^{\alpha\,m_3}-q^{\alpha\,m_2}\Big).
\end{eqnarray*}
\end{small}
\begin{proposition}
	The operators ${\mathbb T}^{q^{\alpha}}_m$ are given by:
	\begin{eqnarray*}\label{tomAC}
	{\mathbb T}^{q^{\alpha}}_m=[m+1+\gamma]_{q^{\alpha}}\,m!\, q^{-\alpha\,m}\,{\partial\over \partial t_m} + {q^{-\alpha(m+1+\gamma)}\over q^{\alpha} - q^{-\alpha}}\sum_{k=1}^{\infty}{(k+m)!\over k!}B_k(t^{\alpha}_1,\cdots,t^{\alpha}_k){\partial\over \partial t_{k+m}}
	\end{eqnarray*}
	and satisfy the product relation
	\begin{eqnarray*}\label{crtomAC}
	&&{\mathbb T}^{q^{\alpha}}_m.{\mathbb T}^{q^{\beta}}_n\sim \,{(q^{\alpha+\beta}-q^{-\alpha-\beta})q^{-m\,\beta}\over q^{\alpha\,m+\beta\,n}(q^{\alpha}-q^{-\alpha})(q^{\beta}-q^{-\beta})}\,{\mathbb T}^{q^{\alpha+\beta}}_{m+n} - {q^{-(n+1)\beta}\over q^{\alpha\,m+\beta\,n}(q^{\beta}-q^{-\beta})}\, {\mathbb T}^{q^{\alpha}}_{m+n}\cr&&\qquad\qquad\qquad\qquad\qquad\qquad\qquad\qquad\qquad\qquad\qquad\qquad - {q^{-m\,\beta}\,q^{-(m+n+1)\alpha}\over q^{\alpha\,m+\beta\,n}(q^{\alpha}-q^{-\alpha})}\,{\mathcal T}^{q^{\beta}}_{m+n}.
	\end{eqnarray*}
\end{proposition}
\subsubsection{Witt $n-$ algebra corresponding to the Jagannathan-Srinivassa deformation \cite{JS}}
	The Witt $n-$ algebra and properties corresponding to {\bf Jagannathan} and {\bf Srinivassa} deformation \cite{JS}    can be obtained by taking ${\mathcal R}(p,q)=1.$ We consider the operators defined as:
\begin{eqnarray}\label{topq}
{\mathcal T}^{p^{\alpha},q^{\alpha}}_m:=-{\mathcal D}_{p^{\alpha},q^{\alpha}}\,z^{m+1},
\end{eqnarray}
where ${\mathcal D}_{p^{\alpha},q^{\alpha}}$ is  the $(p,q)-$ deformed derivative:
\begin{eqnarray*}
	{\mathcal D}_{p^{\alpha},q^{\alpha}}\big(\phi(z)\big)={\phi(p^{\alpha}z)-\phi(q^{\alpha}z)\over p^{\alpha}-q^{\alpha}}.
\end{eqnarray*}
From  the $(p,q)-$ number
(\ref{JSn}), the operators  (\ref{topq}) take the form
\begin{eqnarray*}
	{\mathcal T}^{p^{\alpha},q^{\alpha}}_m=-[m+1]_{p^{\alpha},q^{\alpha}}\,z^{m}.
\end{eqnarray*}
	The operators (\ref{topq}) satisfy  
	\begin{small}
	\begin{eqnarray*}
	 {\mathcal T}^{p^{\alpha},q^{\alpha}}_m.{\mathcal T}^{p^{\beta},q^{\beta}}_n=-{(p^{\alpha+\beta}-q^{\alpha+\beta})p^{-m\beta}\over (p^{\alpha}-q^{\alpha})(p^{\beta}-q^{\beta})}{\mathcal T}^{p^{\alpha+\beta},q^{\alpha+\beta}}_{m+n} + {q^{(n+1)\beta}\over p^{\beta}-q^{\beta}}{\mathcal T}^{p^{\alpha},q^{\alpha}}_{m+n} + {p^{-m\,\beta}q^{(m+n+1)\alpha}\over p^{\alpha}-q^{\alpha}}{\mathcal T}^{p^{\beta},q^{\beta}}_{m+n}
	\end{eqnarray*}
	and the commutation relation
	\begin{eqnarray}\label{crtopq}
	&&\Big[{\mathcal T}^{p^{\alpha},q^{\alpha}}_m, {\mathcal T}^{p^{\beta},q^{\beta}}_n\Big]={(p^{\alpha+\beta}-q^{\alpha+\beta})(p^{-n\,\alpha}-p^{-m\,\beta})\over (p^{\alpha}-q^{\alpha})(p^{\beta}-q^{\beta})}{\mathcal T}^{p^{\alpha+\beta},q^{\alpha+\beta}}_{m+n}-{q^{(m+n+1)\beta}(p^{-n\alpha}-q^{-m\beta})\over p^{\beta}-q^{\beta}}{\mathcal T}^{p^{\alpha},q^{\alpha}}_{m+n}\cr&&\qquad\qquad\qquad\qquad\qquad\qquad\qquad\qquad\qquad\qquad\qquad\qquad  + {q^{(m+n+1)\alpha}(p^{-m\beta}-q^{-n\alpha})\over p^{\alpha}-q^{\alpha}}{\mathcal T}^{p^{\beta},q^{\beta}}_{m+n}.
	\end{eqnarray}
\end{small}
Taking $\alpha=\beta=1$ in the above relation (\ref{crtopq}), we have:
	\begin{eqnarray*}
		\Big[{\mathcal T}^{p,q}_m, {\mathcal T}^{p,q}_n\Big]={(p^{-n}-p^{-m})\over (p-q)}\,[2]_{p,q}{\mathcal T}^{p^{2},q^{2}}_{m+n}-{q^{m+n+1}\over p-q}\Big((p^{-n}-q^{-m})-(p^{-m}-q^{-n})\Big) {\mathcal T}^{p,q}_{m+n}.
	\end{eqnarray*}
The $n-$ bracket is  defined by:
\begin{eqnarray*}
	\Big[{\mathcal T}^{p^{\alpha_1},q^{\alpha_1}}_{m_1},\cdots,{\mathcal T}^{p^{\alpha_n},q^{\alpha_n}}_{m_n}
	\Big]:=\Gamma^{i_1 \cdots i_n}_{1 \cdots n}\,{\mathcal T}^{p^{\alpha_{i_1}},q^{\alpha_{i_1}}}_{m_{i_1}} \cdots {\mathcal T}^{p^{\alpha_{i_n}},q^{\alpha_{i_n}}}_{m_{i_n}}
\end{eqnarray*}
where $\Gamma^{i_1 \cdots i_n}_{1 \cdots n}$ is the L\'evi-Civit\'a symbol given by:
\begin{eqnarray*}
	\Gamma^{j_1 \cdots j_p}_{i_1 \cdots i_p}= det\left( \begin{array} {ccc}
		\delta^{j_1}_{i_1} &\cdots&  \delta^{j_1}_{i_p}   \\ 
		\vdots && \vdots \\
		\delta^{j_p}_{i_1} & \cdots& \delta^{j_p}_{i_p}
		\end {array} \right) .
	\end{eqnarray*}
	The $n-$ bracket  with the same $(p^{\alpha},q^{\alpha})$ is deduced as: 
	\begin{eqnarray*}
		\Big[{\mathcal T}^{p^{\alpha},q^{\alpha}}_{m_1},\cdots,{\mathcal T}^{p^{\alpha},q^{\alpha}}_{m_n}
		\Big]=\Gamma^{1\cdots n}_{1\cdots n}\,{\mathcal T}^{p^{\alpha},q^{\alpha}}_{m_{1}}\cdots {\mathcal T}^{p^{\alpha},q^{\alpha}}_{m_{n}}.
		\end{eqnarray*}
	Putting $\alpha=\beta$ in the relation (\ref{crtopq}), we obtain:
	\begin{eqnarray*}\label{crtobpq}
		&&\Big[{\mathcal T}^{p^{\alpha},q^{\alpha}}_m, {\mathcal T}^{p^{\alpha},q^{\alpha}}_n\Big]={(p^{-n\alpha}-p^{-m\alpha})\over (p^{\alpha}-q^{\alpha})}[2]_{p^{\alpha},q^{\alpha}}{\mathcal T}^{p^{2\alpha},q^{2\alpha}}_{m+n}\cr&&\qquad\qquad\qquad\qquad\qquad\qquad-{q^{(m+n+1)\alpha}\over p^{\alpha}-q^{\alpha}}\Big((p^{-n\alpha}-p^{-m\alpha})+(q^{-n\alpha}-q^{-m\alpha})\Big) {\mathcal T}^{p^{\alpha},q^{\alpha}}_{m+n}
	\end{eqnarray*}
	and the $n-$ bracket is rewritten as follows:
	\begin{eqnarray}\label{crnaJS}
	&&\Big[{\mathcal T}^{p^{\alpha},q^{\alpha}}_{m_1},\cdots, {\mathcal T}^{p^{\alpha},q^{\alpha}}_{m_n}\Big]={(-1)^{n+1}\over (p^{\alpha}-q^{\alpha})^{n-1}}\Big( H^n_{\alpha}\,[n]_{p^{\alpha},q^{\alpha}}{\mathcal T}^{p^{n\alpha},q^{n\alpha}}_{m_1+\cdots+m_n} - [n-1]_{p^{\alpha},q^{\alpha}}\cr&&\qquad\qquad\qquad\qquad\qquad\qquad\qquad\qquad \times q^{\alpha\big(\sum_{l=1}^{n}m_l+1\big)}\big(H^n_{\alpha}+ M^n_{\alpha}\big){\mathcal T}^{p^{(n-1)\alpha},q^{(n-1)\alpha}}_{m_1+\cdots+m_n}\Big),
	\end{eqnarray}
	where 
	\begin{small}
		\begin{eqnarray*}
			&&H^n_{\alpha}= (p^{\alpha}-q^{\alpha})^{n\choose 2}p^{-\alpha(n-1)\sum_{s=1}^{n}m_s}\,\prod_{1\leq j < k \leq n}\Big([m_k]_{p^{\alpha},q^{\alpha}}-[m_j]_{p^{\alpha},q^{\alpha}}\Big)\cr&&\qquad\qquad\qquad\qquad\qquad\qquad +p^{-\alpha(n-1)\sum_{s=1}^{n}m_s}\prod_{1\leq j < k \leq n} \Big(q^{\alpha\,m_k}-q^{\alpha\,m_j}\Big) 
		\end{eqnarray*}
		and 
		\begin{eqnarray*}
		&&	M^{n}_{\alpha}
			=(p^{\alpha}-q^{\alpha})^{n\choose 2}q^{-\alpha(n-1)\sum_{s=1}^{n}m_s}\prod_{1\leq j < k \leq n}\Big([m_k]_{p^{\alpha},q^{\alpha}}-[m_j]_{p^{\alpha},q^{\alpha}}\Big)\cr&&\qquad\qquad\qquad\qquad\qquad\qquad  +(-1)^{n-1}q^{-\alpha(n-1)\sum_{s=1}^{n}m_s}\prod_{1\leq j < k \leq n} \Big(p^{\alpha\,m_k}-p^{\alpha\,m_j}\Big) .
		\end{eqnarray*}
	\end{small}
	Taking $n=3$ in the relation (\ref{crnaJS}), we obtain the $(p,q)-$ Witt $3-$ algebra:
	\begin{eqnarray*}
		&&\Big[{\mathcal T}^{p^{\alpha},q^{\alpha}}_{m_1},{\mathcal T}^{p^{\alpha},q^{\alpha}}_{m_2}, {\mathcal T}^{p^{\alpha},q^{\alpha}}_{m_3}\Big]={1\over (p^{\alpha}-q^{\alpha})^{2}}\Big( H^3_{\alpha}\,[3]_{p^{\alpha},q^{\alpha}}{\mathcal T}^{p^{3\alpha},q^{3\alpha}}_{m_1+\cdots+m_3} - [n-1]_{p^{\alpha},q^{\alpha}}\cr&&\qquad\qquad\qquad\qquad\qquad\qquad\qquad\qquad\qquad\quad \times q^{\alpha\big(\sum_{l=1}^{3}m_l+1\big)}\big(H^3_{\alpha}+ M^3_{\alpha}\big){\mathcal T}^{p^{2\alpha},q^{2\alpha}}_{m_1+\cdots+m_3}\Big),
	\end{eqnarray*}
where
	\begin{small}
		\begin{eqnarray*}
			&&H^3_{\alpha}= (p^{\alpha}-q^{\alpha})^{3\choose 2}p^{-2\alpha\sum_{s=1}^{3}m_s}\,\prod_{1\leq j < k \leq 3}\Big([m_k]_{p^{\alpha},q^{\alpha}}-[m_j]_{p^{\alpha},q^{\alpha}}\Big)\cr&&\qquad\qquad\qquad\qquad\qquad\qquad\qquad\qquad\qquad\qquad +p^{-2\alpha\sum_{s=1}^{3}m_s}\prod_{1\leq j < k \leq 3} \Big(q^{\alpha\,m_k}-q^{\alpha\,m_j}\Big) 
		\end{eqnarray*}
		and 
		\begin{eqnarray*}
			&&	M^{3}_{\alpha}
			=(p^{\alpha}-q^{\alpha})^{3\choose 2}q^{-2\alpha\sum_{s=1}^{3}m_s}\prod_{1\leq j < k \leq 3}\Big([m_k]_{p^{\alpha},q^{\alpha}}-[m_j]_{p^{\alpha},q^{\alpha}}\Big)\cr&&\qquad\qquad\qquad\qquad\qquad\qquad\qquad\qquad\qquad\qquad  +q^{-2\alpha\sum_{s=1}^{3}m_s}\prod_{1\leq j < k \leq 3} \Big(p^{\alpha\,m_k}-p^{\alpha\,m_j}\Big) .
		\end{eqnarray*}
	\end{small}
\begin{proposition}
	The operators ${\mathbb T}^{p^{\alpha},q^{\alpha}}_m$ are given by
	\begin{eqnarray*}\label{tomJS}
		{\mathbb T}^{p^{\alpha},q^{\alpha}}_m=[m+1+\gamma]_{p^{\alpha},q^{\alpha}}m!\, p^{-\alpha\,m}{\partial\over \partial t_m} + {q^{\alpha(m+1+\gamma)}\over p^{\alpha} - q^{\alpha}}\sum_{k=1}^{\infty}{(k+m)!\over k!}B_k(t^{\alpha}_1,\cdots,t^{\alpha}_k){\partial\over \partial t_{k+m}}
	\end{eqnarray*}
	and satisfy the product relation:
	\begin{eqnarray*}\label{crtomJS}
		&&{\mathbb T}^{p^{\alpha},q^{\alpha}}_m.{\mathbb T}^{p^{\beta},q^{\beta}}_n\sim \,{\big(p^{\alpha+\beta}_-q^{\alpha+\beta}\big)p^{-m\,\beta}\over p^{\alpha\,m+\beta\,n}\big(p^{\alpha}-q^{\alpha}\big)\big(p^{\beta}-q^{\beta}\big)}\,{\mathbb T}^{p^{\alpha+\beta},q^{\alpha+\beta}}_{m+n} - {q^{(n+1)\beta}\over p^{\alpha\,m+\beta\,n}\big(p^{\beta}-q^{\beta}\big)}\, {\mathbb T}^{p^{\alpha},q^{\alpha}}_{m+n}\cr&&\qquad\qquad\qquad\qquad\qquad\qquad\qquad\qquad\qquad\qquad\qquad\qquad - {p^{-m\,\beta}\,q^{(m+n+1)\alpha}\over p^{\alpha\,m+\beta\,n}\big(p^{\alpha}-q^{\alpha}\big)}\,{\mathcal T}^{p^{\beta},q^{\beta}}_{m+n}.
	\end{eqnarray*}
\end{proposition}
\subsubsection{Witt $n-$ algebra associated to the Chakrabarty and Jagannathan deformation \cite{CJ}}
	Taking $\epsilon_1=p^{-1}$ and $\epsilon_2=q,$ we obtain the $(p^{-1},q)-$ deformed Witt $n-$ algebra. The $(p^{-1},q)-$ deformed derivative is defined by
\begin{eqnarray*}
	{\mathcal D}_{p^{-\alpha},q^{\alpha}}\big(\phi(z)\big):={\phi(p^{-\alpha}z)-\phi(q^{\alpha}z)\over p^{-\alpha}-q^{\alpha}}
\end{eqnarray*}
and 
the operators ${\mathcal T}^{p^{-\alpha},q^{\alpha}}_m$ by:
\begin{eqnarray}\label{toCJ}
{\mathcal T}^{p^{-\alpha},q^{\alpha}}_m:=-{\mathcal D}_{p^{-\alpha},q^{\alpha}}\,z^{m+1}.
\end{eqnarray}
From  the $(p^{-1},q)-$ number
(\ref{JSn}), the operators  (\ref{toCJ}) take the form
\begin{eqnarray*}
	{\mathcal T}^{p^{-\alpha},q^{\alpha}}_m=-[m+1]_{p^{-\alpha},q^{\alpha}}\,z^{m}.
\end{eqnarray*}
The operators (\ref{toCJ}) satisfy  
\begin{small}
	\begin{eqnarray*}
	&&{\mathcal T}^{p^{-\alpha},q^{\alpha}}_m.{\mathcal T}^{p^{-\beta},q^{\beta}}_n=-{(p^{-\alpha-\beta}-q^{\alpha+\beta})p^{m\beta}\over (p^{-\alpha}-q^{\alpha})(p^{-\beta}-q^{\beta})}{\mathcal T}^{p^{-\alpha-\beta},q^{\alpha+\beta}}_{m+n} + {q^{(n+1)\beta}\over p^{-\beta}-q^{\beta}}{\mathcal T}^{p^{-\alpha},q^{\alpha}}_{m+n}\cr&&\qquad\qquad\qquad\qquad\qquad\qquad\qquad\qquad\qquad\qquad\qquad\qquad\qquad\qquad + {p^{m\,\beta}q^{(m+n+1)\alpha}\over p^{-\alpha}-q^{\alpha}}{\mathcal T}^{p^{-\beta},q^{\beta}}_{m+n}
	\end{eqnarray*}
	and the commutation relation
	\begin{eqnarray}\label{crtoCJ}
	&&\Big[{\mathcal T}^{p^{-\alpha},q^{\alpha}}_m, {\mathcal T}^{p^{-\beta},q^{\beta}}_n\Big]={(p^{-\alpha-\beta}-p^{-\alpha-\beta})(p^{n\alpha}-p^{m\beta})\over (p^{-\alpha}-q^{\alpha})(p^{-\beta}-q^{\beta})}{\mathcal T}^{p^{-\alpha-\beta},q^{\alpha+\beta}}_{m+n}-{q^{(m+n+1)\beta}(p^{n\alpha}-q^{-m\beta})\over p^{-\beta}-q^{\beta}}{\mathcal T}^{p^{-\alpha},q^{\alpha}}_{m+n}\cr&&\qquad\qquad\qquad\qquad\qquad\qquad\qquad\qquad\qquad\qquad\qquad\qquad  + {q^{(m+n+1)\alpha}(p^{m\beta}-q^{-n\alpha})\over p^{-\alpha}-q^{\alpha}}{\mathcal T}^{p^{-\beta},q^{\beta}}_{m+n}.
	\end{eqnarray}
\end{small}
Taking $\alpha=\beta=1$ in the above relation (\ref{crtoCJ}), we have:
\begin{eqnarray*}
	\Big[{\mathcal T}^{p^{-1},q}_m, {\mathcal T}^{p^{-1},q}_n\Big]={(p^{n}-p^{m})\over (p^{-1}-q)}[2]_{p^{-1},q}{\mathcal T}^{p^{-2},q^{2}}_{m+n}-{q^{m+n+1}\over p^{-1}-q}\Big((p^{n}-q^{-m})-(p^{m}-q^{-n})\Big) {\mathcal T}^{p^{-1},q}_{m+n}.
\end{eqnarray*}
We consider the $n-$ bracket defined by:
\begin{eqnarray*}
	\Big[{\mathcal T}^{p^{-\alpha_1},q^{\alpha_1}}_{m_1},\cdots,{\mathcal T}^{p^{-\alpha_n},q^{\alpha_n}}_{m_n}
	\Big]:=\Gamma^{i_1 \cdots i_n}_{1 \cdots n}\,{\mathcal T}^{p^{-\alpha_{i_1}},q^{\alpha_{i_1}}}_{m_{i_1}} \cdots {\mathcal T}^{p^{-\alpha_{i_n}},q^{\alpha_{i_n}}}_{m_{i_n}}
\end{eqnarray*}
where $\Gamma^{i_1 \cdots i_n}_{1 \cdots n}$ is the L\'evi-Civit\'a symbol given by:
\begin{eqnarray*}
	\Gamma^{j_1 \cdots j_p}_{i_1 \cdots i_p}= det\left( \begin{array} {ccc}
		\delta^{j_1}_{i_1} &\cdots&  \delta^{j_1}_{i_p}   \\ 
		\vdots && \vdots \\
		\delta^{j_p}_{i_1} & \cdots& \delta^{j_p}_{i_p}
		\end {array} \right) .
	\end{eqnarray*}
	Our study is focussed on the case with the same $(p^{-\alpha},q^{\alpha}).$ Thus, we have
	\begin{eqnarray*}
		\Big[{\mathcal T}^{p^{-\alpha},q^{\alpha}}_{m_1},\cdots,{\mathcal T}^{p^{-\alpha},q^{\alpha}}_{m_n}
		\Big]=\Gamma^{1\cdots n}_{1\cdots n}\,{\mathcal T}^{p^{-\alpha},q^{\alpha}}_{m_{1}}\cdots {\mathcal T}^{p^{-\alpha},q^{\alpha}}_{m_{n}}.
	\end{eqnarray*}
	Putting $\alpha=\beta$ in the relation (\ref{crtoCJ}), we obtain:
	\begin{eqnarray*}\label{crtobCJ}
	&&\Big[{\mathcal T}^{p^{-\alpha},q^{\alpha}}_m, {\mathcal T}^{p^{-\alpha},q^{\alpha}}_n\Big]={(p^{n\alpha}-p^{m\alpha})\over (p^{-\alpha}-q^{\alpha})}[2]_{p^{-\alpha},q^{\alpha}}{\mathcal T}^{p^{-2\alpha},q^{2\alpha}}_{m+n}\cr&&\qquad\qquad\qquad\qquad\qquad\qquad-{q^{(m+n+1)\alpha}\over p^{-\alpha}-q^{\alpha}}\Big((p^{n\alpha}-p^{m\alpha})+(q^{-n\alpha}-q^{-m\alpha})\Big) {\mathcal T}^{p^{-\alpha},q^{\alpha}}_{m+n}.
	\end{eqnarray*}
	Then the $n-$ bracket is given by:
	\begin{eqnarray}\label{crnaCJ}
		&&\Big[{\mathcal T}^{p^{-\alpha},q^{\alpha}}_{m_1},\cdots, {\mathcal T}^{p^{-\alpha},q^{\alpha}}_{m_n}\Big]={(-1)^{n+1}\over (p^{-\alpha}-q^{\alpha})^{n-1}}\Big( H^n_{\alpha}\,[n]_{p^{-\alpha},q^{\alpha}}{\mathcal T}^{p^{-n\alpha},q^{n\alpha}}_{m_1+\cdots+m_n} - [n-1]_{p^{-\alpha},q^{\alpha}}\cr&&\qquad\qquad\qquad\qquad\qquad\qquad\qquad\qquad \times q^{\alpha\big(\sum_{l=1}^{n}m_l+1\big)}\big(H^n_{\alpha}+ M^n_{\alpha}\big){\mathcal T}^{p^{-(n-1)\alpha},q^{(n-1)\alpha}}_{m_1+\cdots+m_n}\Big),
	\end{eqnarray}
	where 
	\begin{small}
		\begin{eqnarray*}
			&&H^n_{\alpha}= (p^{-\alpha}-q^{\alpha})^{n\choose 2}p^{\alpha(n-1)\sum_{s=1}^{n}m_s}\,\prod_{1\leq j < k \leq n}\Big([m_k]_{p^{-\alpha},q^{\alpha}}-[m_j]_{p^{-\alpha},q^{\alpha}}\Big)\cr&&\qquad\qquad\qquad\qquad\qquad\qquad +p^{\alpha(n-1)\sum_{s=1}^{n}m_s}\prod_{1\leq j < k \leq n} \Big(q^{\alpha\,m_k}-q^{\alpha\,m_j}\Big) 
		\end{eqnarray*}
		and 
		\begin{eqnarray*}
			&&M^{n}_{\alpha}
			=(p^{-\alpha}-q^{\alpha})^{n\choose 2}q^{-\alpha(n-1)\sum_{s=1}^{n}m_s}_2\prod_{1\leq j < k \leq n}\Big([m_k]_{p^{-\alpha},q^{\alpha}}-[m_j]_{p^{-\alpha},q^{\alpha}}\Big)\cr&&\qquad\qquad\qquad\qquad\qquad\qquad  +(-1)^{n-1}q^{-\alpha(n-1)\sum_{s=1}^{n}m_s}\prod_{1\leq j < k \leq n} \Big(p^{-\alpha\,m_k}-p^{-\alpha\,\_j}\Big).
		\end{eqnarray*}
	\end{small}
	Setting $n=3$ in the relation (\ref{crnaCJ}), we obtain the $(p,q)-$ Witt $3-$ algebra:
	\begin{eqnarray*}
		&&\Big[{\mathcal T}^{p^{-\alpha},q^{\alpha}}_{m_1},{\mathcal T}^{p^{-\alpha},q^{\alpha}}_{m_2}, {\mathcal T}^{p^{-\alpha},q^{\alpha}}_{m_3}\Big]={1\over (p^{-\alpha}-q^{\alpha})^{2}}\Big( H^3_{\alpha}\,[3]_{p^{-\alpha},q^{\alpha}}{\mathcal T}^{p^{-3\alpha},q^{3\alpha}}_{m_1+\cdots+m_3} - [n-1]_{p^{-\alpha},q^{\alpha}}\cr&&\qquad\qquad\qquad\qquad\qquad\qquad\qquad\qquad\qquad\quad \times q^{\alpha\big(\sum_{l=1}^{3}m_l+1\big)}\big(H^3_{\alpha}+ M^3_{\alpha}\big){\mathcal T}^{p^{-2\alpha},q^{2\alpha}}_{m_1+\cdots+m_3}\Big),
	\end{eqnarray*}
where
	\begin{small}
		\begin{eqnarray*}
			&&H^3_{\alpha}= (p^{-\alpha}-q^{\alpha})^{3\choose 2}p^{2\alpha\sum_{s=1}^{3}m_s}\,\prod_{1\leq j < k \leq 3}\Big([m_k]_{p^{-\alpha},q^{\alpha}}-[m_j]_{p^{-\alpha},q^{\alpha}}\Big)\cr&&\qquad\qquad\qquad\qquad\qquad\qquad\qquad\qquad\qquad +p^{2\alpha\sum_{s=1}^{3}m_s}\prod_{1\leq j < k \leq 3} \Big(q^{\alpha\,m_k}-q^{\alpha\,m_j}\Big) 
		\end{eqnarray*}
		and 
		\begin{eqnarray*}
			&&M^{3}_{\alpha}
			=(p^{-\alpha}-q^{\alpha})^{3\choose 2}q^{-2\alpha\sum_{s=1}^{3}m_s}_2\prod_{1\leq j < k \leq 3}\Big([m_k]_{p^{-\alpha},q^{\alpha}}-[m_j]_{p^{-\alpha},q^{\alpha}}\Big)\cr&&\qquad\qquad\qquad\qquad\qquad\qquad\qquad\qquad\qquad + q^{-2\alpha\sum_{s=1}^{3}m_s}\prod_{1\leq j < k \leq 3} \Big(p^{-\alpha\,m_k}-p^{-\alpha\,m_j}\Big).
		\end{eqnarray*}
	\end{small}
\begin{proposition}
	The operators ${\mathbb T}^{p^{-\alpha},q^{\alpha}}_m$ are given by
	\begin{small}
		\begin{eqnarray*}\label{tomCJ}
			{\mathbb T}^{p^{-\alpha},q^{\alpha}}_m=[m+1+\gamma]_{p^{-\alpha},q^{\alpha}}m!\, p^{\alpha\,m}{\partial\over \partial t_m} + {q^{\alpha(m+1+\gamma)}\over p^{-\alpha} - q^{\alpha}}\sum_{k=1}^{\infty}{(k+m)!\over k!}B_k(t^{\alpha}_1,\cdots,t^{\alpha}_k){\partial\over \partial t_{k+m}}
		\end{eqnarray*}
		and satisfy the product relation:
		\begin{eqnarray*}\label{crtomCJ}
			&&{\mathbb T}^{p^{-\alpha},q^{\alpha}}_m.{\mathbb T}^{p^{-\beta},q^{\beta}}_n\sim \,{\big(p^{-\alpha-\beta}_-q^{\alpha+\beta}\big)p^{-m\,\beta}\over p^{-\alpha\,m-\beta\,n}\big(p^{-\alpha}-q^{\alpha}\big)\big(p^{-\beta}-q^{\beta}\big)}\,{\mathbb T}^{p^{-\alpha-\beta},q^{\alpha+\beta}}_{m+n} - {q^{(n+1)\beta}\over p^{-\alpha\,m-\beta\,n}\big(p^{-\beta}-q^{\beta}\big)}\, {\mathbb T}^{p^{-\alpha},q^{\alpha}}_{m+n}\cr&&\qquad\qquad\qquad\qquad\qquad\qquad\qquad\qquad\qquad\qquad\qquad\qquad - {p^{-m\,\beta}\,q^{(m+n+1)\alpha}\over p^{-\alpha\,m-\beta\,n}\big(p^{-\alpha}-q^{\alpha}\big)}\,{\mathcal T}^{p^{-\beta},q^{\beta}}_{m+n}.
		\end{eqnarray*}
	\end{small}
\end{proposition}
\subsubsection{Witt $n-$ algebra induced by the Hounkonnou-Ngompe generalization of $q-$ Quesne deformation \cite{HN}}
	The Witt  $n-$ algebra and properties are obtained by taking $\epsilon_1=p$ and $\epsilon_2=q^{-1}.$
we define the operators ${\mathcal T}^{p^{\alpha},q^{\alpha}}_m$ as follows: 
\begin{eqnarray}\label{toHN}
{\mathcal T}^{p^{\alpha},q^{\alpha}}_m:=-{\mathcal D}_{p^{\alpha},q^{\alpha}}\,z^{m+1},
\end{eqnarray}
where ${\mathcal D}_{p^{\alpha},q^{\alpha}}$ is the derivative 
\begin{eqnarray*}
	{\mathcal D}_{p^{\alpha},q^{\alpha}}\big(\phi(z)\big)={\phi(p^{\alpha}z)-\phi(q^{-\alpha}z)\over q^{-\alpha}-p^{\alpha}}.
\end{eqnarray*}
The operators (\ref{toHN}) satisfy  
\begin{small}
	\begin{eqnarray*}
	&&{\mathcal T}^{p^{\alpha},q^{\alpha}}_m.{\mathcal T}^{p^{\beta},q^{\beta}}_n=-{(p^{\alpha+\beta}-q^{-\alpha-\beta})p^{-m\beta}\over (p^{\alpha}-q^{-\alpha})(p^{\beta}-q^{-\beta})}{\mathcal T}^{p^{\alpha+\beta},q^{\alpha+\beta}}_{m+n} + {q^{-(n+1)\beta}\over p^{\beta}-q^{-\beta}}{\mathcal T}^{p^{\alpha},q^{\alpha}}_{m+n}\cr&&\qquad\qquad\qquad\qquad\qquad\qquad\qquad\qquad\qquad\qquad\qquad\qquad\qquad + {p^{-m\,\beta}q^{-(m+n+1)\alpha}\over p^{\alpha}-q^{-\alpha}}{\mathcal T}^{p^{\beta},q^{\beta}}_{m+n}
	\end{eqnarray*}
	and the commutation relation
	\begin{eqnarray}\label{crtoHN}
	&&\Big[{\mathcal T}^{p^{\alpha},q^{\alpha}}_m, {\mathcal T}^{p^{\beta},q^{\beta}}_n\Big]={(p^{\alpha+\beta}-q^{-\alpha-\beta})(p^{-n\,\alpha}-p^{-m\,\beta})\over (p^{\alpha}-q^{-\alpha})(p^{\beta}-q^{-\beta})}{\mathcal T}^{p^{\alpha+\beta},q^{\alpha+\beta}}_{m+n}-{q^{-(m+n+1)\beta}(p^{-n\alpha}-q^{m\beta})\over p^{\beta}-q^{-\beta}}{\mathcal T}^{p^{\alpha},q^{\alpha}}_{m+n}\cr&&\qquad\qquad\qquad\qquad\qquad\qquad\qquad\qquad\qquad\qquad\qquad\qquad  + {q^{(m+n+1)\alpha}(p^{-m\beta}-q^{n\alpha})\over p^{\alpha}-q^{-\alpha}}{\mathcal T}^{p^{\beta},q^{\beta}}_{m+n}.
	\end{eqnarray}
\end{small}
Putting $\alpha=\beta=1$ in  relation (\ref{crtoHN}), yields:
\begin{eqnarray*}
	\Big[{\mathcal T}^{p,q}_m, {\mathcal T}^{p,q}_n\Big]={q(p^{-n}-p^{-m})\over p(p-q^{-1})}\,[2]_{p,q}{\mathcal T}^{p^{2},q^{2}}_{m+n}-{q^{-(m+n+1)}\over p-q^{-1}}\Big((p^{-n}-q^{m})-(p^{-m}-q^{n})\Big) {\mathcal T}^{p,q}_{m+n}.
\end{eqnarray*}
We consider the $n-$ bracket defined by:
\begin{eqnarray*}
	\Big[{\mathcal T}^{p^{\alpha_1},q^{\alpha_1}}_{m_1},\cdots,{\mathcal T}^{p^{\alpha_n},q^{\alpha_n}}_{m_n}
	\Big]:=\Gamma^{i_1 \cdots i_n}_{1 \cdots n}\,{\mathcal T}^{p^{\alpha_{i_1}},q^{\alpha_{i_1}}}_{m_{i_1}} \cdots {\mathcal T}^{p^{\alpha_{i_n}},q^{\alpha_{i_n}}}_{m_{i_n}}
\end{eqnarray*}
where $\Gamma^{i_1 \cdots i_n}_{1 \cdots n}$ is the L\'evi-Civit\'a symbol given by:
\begin{eqnarray*}
	\Gamma^{j_1 \cdots j_p}_{i_1 \cdots i_p}= det\left( \begin{array} {ccc}
		\delta^{j_1}_{i_1} &\cdots&  \delta^{j_1}_{i_p}   \\ 
		\vdots && \vdots \\
		\delta^{j_p}_{i_1} & \cdots& \delta^{j_p}_{i_p}
		\end {array} \right) .
	\end{eqnarray*}
	Our study is focussed on the case with the same $(p^{\alpha},q^{\alpha}).$ Thus, we have
	\begin{eqnarray*}
		\Big[{\mathcal T}^{p^{\alpha},q^{\alpha}}_{m_1},\cdots,{\mathcal T}^{p^{\alpha},q^{\alpha}}_{m_n}
		\Big]=\Gamma^{1\cdots n}_{1\cdots n}\,{\mathcal T}^{p^{\alpha},q^{\alpha}}_{m_{1}}\cdots {\mathcal T}^{p^{\alpha},q^{\alpha}}_{m_{n}}.
	\end{eqnarray*}
	Putting $\alpha=\beta$ in the relation (\ref{crtoHN}), we obtain:
	\begin{eqnarray}\label{crtobHN}
	&&\Big[{\mathcal T}^{p^{\alpha},q^{\alpha}}_m, {\mathcal T}^{p^{\alpha},q^{\alpha}}_n\Big]={q^{\alpha}(p^{-n\alpha}-p^{-m\alpha})\over p^{\alpha}(p^{\alpha}-q^{-\alpha})}[2]_{p^{\alpha},q^{\alpha}}{\mathcal T}^{p^{2\alpha},q^{2\alpha}}_{m+n}\cr&&\qquad\qquad\qquad\qquad\qquad\qquad-{q^{-(m+n+1)\alpha}\over p^{\alpha}-q^{-\alpha}}\Big((p^{-n\alpha}-p^{-m\alpha})+(q^{n\alpha}-q^{m\alpha})\Big) {\mathcal T}^{p^{\alpha},q^{\alpha}}_{m+n}.
	\end{eqnarray}
	From the above relation (\ref{crtobHN}), we deduce the $n-$ bracket as follows:
	\begin{eqnarray}\label{crnaHN}
		&&\Big[{\mathcal T}^{p^{\alpha},q^{\alpha}}_{m_1},\cdots, {\mathcal T}^{p^{\alpha},q^{\alpha}}_{m_n}\Big]={(-1)^{n+1}\big(qp^{-1}\big)^{\alpha}\over (p^{\alpha}-q^{-\alpha})^{n-1}}\Big( H^n_{\alpha}[n]_{p^{\alpha},q^{\alpha}}{\mathcal T}^{p^{n\alpha},q^{n\alpha}}_{m_1+\cdots+m_n} - [n-1]_{p^{\alpha},q^{\alpha}}\cr&&\qquad\qquad\qquad\qquad\qquad\qquad\qquad\qquad \times q^{-\alpha\big(\sum_{l=1}^{n}m_l+1\big)}\big(H^n_{\alpha}+ M^n_{\alpha}\big){\mathcal T}^{p^{(n-1)\alpha},q^{(n-1)\alpha}}_{m_1+\cdots+m_n}\Big),
	\end{eqnarray}
	where 
	\begin{small}
		\begin{eqnarray*}
			&&H^n_{\alpha}= (p^{\alpha}-q^{-\alpha})^{n\choose 2}p^{-\alpha(n-1)\sum_{s=1}^{n}m_s}\,\prod_{1\leq j < k \leq n}\big(qp^{-1}\big)^{\alpha}\Big([m_k]_{p^{\alpha},q^{\alpha}}-[m_j]_{p^{\alpha},q^{\alpha}}\Big)\cr&&\qquad\qquad\qquad\qquad\qquad\qquad +p^{-\alpha(n-1)\sum_{s=1}^{n}m_s}\prod_{1\leq j < k \leq n} \Big(q^{-\alpha\,m_k}-q^{-\alpha\,m_j}\Big) 
		\end{eqnarray*}
		and 
		\begin{eqnarray*}
			&&M^{n}_{\alpha}
			=(p^{\alpha}-q^{-\alpha})^{n\choose 2}q^{\alpha(n-1)\sum_{s=1}^{n}m_s}\prod_{1\leq j < k \leq n}\big(qp^{-1}\big)^{\alpha}\Big([m_k]_{p^{\alpha},q^{\alpha}}-[m_j]_{p^{\alpha},q^{\alpha}}\Big)\cr&&\qquad\qquad\qquad\qquad\qquad\qquad  +(-1)^{n-1}q^{\alpha(n-1)\sum_{s=1}^{n}m_s}\prod_{1\leq j < k \leq n} \Big(p^{\alpha\,m_k}-p^{\alpha\,m_j}\Big).
		\end{eqnarray*}
	\end{small}
	Taking $n=3$ in the relation (\ref{crnaHN}), we obtain the Witt $3-$ algebra correponding to the {\bf Hounkonnou-Ngompe} generalization of  $q-$ Quesne deformation\cite{HN}:
	\begin{eqnarray*}
		&&\Big[{\mathcal T}^{p^{\alpha},q^{\alpha}}_{m_1},{\mathcal T}^{p^{\alpha},q^{\alpha}}_{m_2}, {\mathcal T}^{p^{\alpha},q^{\alpha}}_{m_3}\Big]={\big(qp^{-1}\big)^{\alpha}\over (p^{\alpha}-q^{-\alpha})^{2}}\Big( H^3_{\alpha}\,[3]_{p^{\alpha},q^{\alpha}}{\mathcal T}^{p^{3\alpha},q^{3\alpha}}_{m_1+\cdots+m_3} - [n-1]_{p^{\alpha},q^{\alpha}}\cr&&\qquad\qquad\qquad\qquad\qquad\qquad\qquad\qquad\qquad\quad \times q^{-\alpha\big(\sum_{l=1}^{3}m_l+1\big)}\big(H^3_{\alpha}+ M^3_{\alpha}\big){\mathcal T}^{p^{2\alpha},q^{2\alpha}}_{m_1+\cdots+m_3}\Big),
	\end{eqnarray*}
	where 
	\begin{small}
		\begin{eqnarray*}
			&&H^3_{\alpha}= (p^{\alpha}-q^{\alpha})^{3\choose 2}p^{-2\alpha\sum_{s=1}^{3}m_s}\,\prod_{1\leq j < k \leq 3}\big(qp^{-1}\big)^{\alpha}\Big([m_k]_{p^{\alpha},q^{\alpha}}-[m_j]_{p^{\alpha},q^{\alpha}}\Big)\cr&&\qquad\qquad\qquad\qquad\qquad\qquad\qquad\qquad\qquad +p^{2\alpha\sum_{s=1}^{3}m_s}\prod_{1\leq j < k \leq 3} \Big(q^{-\alpha\,m_k}-q^{-\alpha\,m_j}\Big) 
		\end{eqnarray*}
		and 
		\begin{eqnarray*}
			&&M^{3}_{\alpha}
			=(p^{\alpha}-q^{-\alpha})^{3\choose 2}q^{2\alpha\sum_{s=1}^{3}m_s}\prod_{1\leq j < k \leq 3}\big(qp^{-1}\big)^{\alpha}\Big([m_k]_{p^{-\alpha},q^{\alpha}}-[m_j]_{p^{\alpha},q^{\alpha}}\Big)\cr&&\qquad\qquad\qquad\qquad\qquad\qquad\qquad\qquad\qquad + q^{-2\alpha\sum_{s=1}^{3}m_s}\prod_{1\leq j < k \leq 3} \Big(p^{\alpha\,m_k}-p^{\alpha\,m_j}\Big).
		\end{eqnarray*}
	\end{small}
\begin{proposition}
	The operators ${\mathbb T}^{p^{\alpha},q^{\alpha}}_m$ are given by
	\begin{small}
		\begin{eqnarray*}\label{tomHN}
			{\mathbb T}^{p^{\alpha},q^{\alpha}}_m={\big(qp^{-1}\big)^{\alpha}\over p^{\alpha\,m}}[m+1+\gamma]_{p^{\alpha},q^{-\alpha}}m!\, {\partial\over \partial t_m} + {q^{-\alpha(m+1+\gamma)}\over p^{\alpha} - q^{-\alpha}}\sum_{k=1}^{\infty}{(k+m)!\over k!}B_k(t^{\alpha}_1,\cdots,t^{\alpha}_k){\partial\over \partial t_{k+m}}
		\end{eqnarray*}
	\end{small}
	and satisfy the product relation:
	\begin{eqnarray*}\label{crtomHN}
		&&{\mathbb T}^{p^{\alpha},q^{\alpha}}_m.{\mathbb T}^{p^{\beta},q^{\beta}}_n\sim \,{\big(p^{\alpha+\beta}-q^{-\alpha-\beta}\big)p^{-m\,\beta}\over p^{\alpha\,m+\beta\,n}\big(p^{\alpha}-q^{-\alpha}\big)\big(p^{\beta}-q^{-\beta}\big)}\,{\mathbb T}^{p^{\alpha+\beta},q^{\alpha+\beta}}_{m+n} - {q^{-(n+1)\beta}\over p^{\alpha\,m+\beta\,n}\big(p^{\beta}-q^{-\beta}\big)}\, {\mathbb T}^{p^{\alpha},q^{\alpha}}_{m+n}\cr&&\qquad\qquad\qquad\qquad\qquad\qquad\qquad\qquad\qquad\qquad\qquad\qquad - {p^{-m\,\beta}\,q^{-(m+n+1)\alpha}\over p^{\alpha\,m+\beta\,n}\big(p^{\alpha}-q^{-\alpha}\big)}\,{\mathcal T}^{p^{\beta},q^{\beta}}_{m+n}.
	\end{eqnarray*}
\end{proposition}
\section{Concluding and remarks}
We have developed a unified framework for deforming Witt and Virasoro algebras from quantum algebras and their generalizations, as well as for establishing deformed quantum Korteweg-de Vries equations. Furthermore, we have extended this study to the construction of the deformed Witt $n-$ algebra, and derived deformed Virasoro constraints for matrix models. Interesting particular quantum deformations have been investigated and discussed. 

\begin{acknowledgements}
This work is supported by TWAS Research Grant RGA No. 17 - 542 RG / MATHS / AF / AC \_G  -FR3240300147. The ICMPA-UNESCO Chair is in partnership with Daniel Iagolnitzer Foundation (DIF), France, and the Association pour la Promotion Scientifique de l'Afrique (APSA), supporting the development of mathematical physics in Africa. Melanija Mitrovi\'c is supported by the Faculty of Mechanical Engineering , University of Ni\v s, Serbia, Grant " Research and development of new generation machine systems in the function of the technological development of Serbia".
\end{acknowledgements}


\begin{thebibliography}{15}
	\bibitem{AS} Aizawa, N. and  Sato, H.: $q-$ deformation of the Virasoro algebra with central extension, {\it Phys. Lett. B.} {\bf 256}, (1991).
	\bibitem{AC} Arik, M. and  Coon, D. D.:   Hilbert spaces of analytic functions and generated
	coherent states, {\it J. Math. Phys.} {\bf 17}, 424-427, (1976).
	\bibitem{BL} Bernard, D. and  LeClair, A.: $q-$deformation of SU(1,1) conformal ward identities and $q-$strings, {\it Phys. Lett. B.} {\bf 227}, 417-423, (1989).
	\bibitem{BPZ} Belavin, A.,  Polyakov, A. and  Zamolodchikov, A.:  Infinite conformal symmetry in two-dimensional quantum field theory, {\it Nucl. Phys. B.} {\bf 241}, 333-380,  (1984).
	\bibitem{CE} Cartan, E.: Les groupes de transformations continus, infinis, simples, {\it Ann. Sci.
		Ecole Norm. Sup.} {\bf 26}, 93-161,  (1909).
	\bibitem{CILPP} Chaichian, M.,  Isaev, A. P., Lukierski, J., Popowicz, Z. and  Presnajder, P.: $q$-deformations of Virasoro algebra and conformal dimensions, {\it Phys. Lett. B.} {\bf 262}, (1991).
	\bibitem{CPPa}  Chaichian, M.,  Popowicz, Z. and  Presnajder, P.:   $q-$ Virasoro algebra and its relation to the $q-$ deformed KdV system, {\it Phys. Lett. B.} {\bf 249}, (1990).
	\bibitem{CP} Chaichian, M. and  Presnajder, P.: Sugawara construction
	and the $q-$ deformation of Virasoro (super) algebra, {\it Phys. Lett. B.} {\bf  277}, 109-118, (1992).
	\bibitem{CJ}  Chakrabarti, R. and  Jagannathan, R.:  A $( p , q)$-deformed Virasoro algebra, {\it J. Phys. A Math. Gen.} {\bf 25}, 2607-2614, (1992). 
	\bibitem{Chakrabarti&Jagan}  Chakrabarti, R. and  Jagannathan, R.: {A $(p, q)-$oscillator realisation of two-parameter quantum algebras,} { J. Phys. \rm A: Math. Gen.} {\bf 24}, L711-L718, (1991).
	\bibitem{CZ}  Curtright, T. and  Zachos, C.: Deforming maps for quantum algebras,   {\it Phys. Lett. B.} {\bf 243}, 237,  (1990).
	\bibitem{FF} Fuchssteiner, B. and  Fokas, A. S.:  
	Symplectic structures, their Bäcklund transformations and hereditary symmetries, {\it Physica. D.} {\bf 4}, 47- 66,  (1981).
	\bibitem{GF} Gelfand, I. M. and  Fuchs, D. B.: Cohomologies of the Lie algebra of vector fields on
	the circle, {\it Funct. Anal. Appl.} {\bf 2}, 342-343, (1968).
	\bibitem{G} Gervais, J. L.: Transport matrices associated with the Virasoro algebra, {\it Phys. Lett. B.}	{\bf 160}, 279-282, (1985).
	\bibitem{Heine} Heine, E.: {\it Handbuch der Kugelfunctionen, Theorie und Anwendungen}, vol.2, G.Reimer,
	Berlin, 1881.
	\bibitem{HB} Hounkonnou, M. N.  and  Kyemba Bukweli, J. D.:  $\mathcal{R}(p,q)-$ calculus: differentiation and integration, {\it SUT Journal of Mathematics}, Vol {\bf 49}  (2), 145-167, (2013). 
	\bibitem{HB1} Hounkonnou, M. N. and  Bukweli Kyemba, J. D.: $\mathcal{R}(p, q)$-deformed
	quantum algebras: coherent states and special functions, {\it J. Math. Phys.} {\bf 51}, 063518, 
	(2010).
	\bibitem{HN}  Hounkonnou, M. N. and  Ngompe Nkouankam, E. B.: On $(p, q, \mu, \nu, \varphi_1, \varphi_2)$ generalized
	oscillator algebra and related bibasic hypergeometric functions, {\it J. Phys.
		A: Math. Theor.} {\bf 40}, 8835-8843, (2007).
	\bibitem{Hounkonnou:2015laa} 
	 Hounkonnou, M. N.,  Guha, P. and  Ratiu, T.:
	Generalized Virasoro algebra: left-symmetry and related algebraic and hydrodynamic properties,
	{\it J. Nonlin. Math. Phys.} vol 23, Iss 1,(2016) .
	\bibitem{HM} Hounkonnou, M. N. and  Melong, F.: $\mathcal{R}(p,q)-$ deformed conformal Virasoro algebra, {\it J. Maths. Phys.}{\bf 60}, (2019).
	\bibitem{HZ} Huang, Q. and  Zhdanov, R.:  Realizations of the Witt and Virasoro algebras and integrable equations, {\it J. Nonlin. Math. Phys.} {\bf 27}, 36-56,   (2019).
	\bibitem{IK} Iohara, K. and  Koga, Y.: {\it Representation Theory of the Virasoro Algebra} (Springer-Verlag, London, 2011).
	\bibitem{JS}Jagannathan, R. and  Srinivasa Rao, K.: Two-parameter quantum algebras,
	twin-basic numbers, and associated generalized hypergeometric series, {\it Proceedings of the International
		Conference on Number Theory and Mathematical Physics}, 20-21 December 2005.
	\bibitem{KB} Kupershmidt, B. A.: On the nature of the Virasoro algebra, {\it J. Nonlin. Math. Phys.} {\bf 6} (2), 222-245, (1998).
	\bibitem{NZ} Nedelin, A. and  Zabzine, M.: $q-$ Virasoro constraints in matrix models, {\it J. High Energy Phys.} {\bf 03}, 098 (2017).
	\bibitem{TN} Nishino, T.:  {\it Function theory in several complex variables}, Translations of mathematical monographs, Volume 193, American Mathematical Society, Providence, Rhode Island, 2001.
	\bibitem{QPT} Quesne, C.,  Penson, K. A. and  Tkachuk, V. M.: Maths-type $q-$deformed coherent
	states for $q > 1,$  {\it Phys. Lett. A.} {\bf 313}, 29-36, (2003).
	\bibitem{RY} Ratindranath, A. and  Yasuhiro, O.: Strings in curved space-time: Virasoro algebra in the classical and quantum theory, {\it Phys. rev. D: Particles and fields.} {\bf 35}, 1917-1938, (1987).
	\bibitem{S} Sato, H.: Realizations of  $q-$ deformed Virasoro algebra, {\it Pro.  Theor. Phys.} {\bf 89}, (1993).
	\bibitem{WYWZ} Wang, R., Yao, Li, M.,  Wu, K. and Zhao, W.: On deformations of the Witt $n-$ algebra, {\it J. Maths. Phys.}{\bf 59}, (2018). 
	\bibitem{WA} Witt, E.: Collected papers. Gesammelte Abhandlungen, Berlin, New York: Springer-Verlag, ISBN 978-3-540-57061-5, MR 1643949, (1998).
	\bibitem{W}  Woronowicz, S. L.: Twisted $SU(2)-$ group:An example of a non-commutative differential calculus, {\it Publ. RIMS, Kyoto University.}  {\bf 23},  117-181, ( 1987).
\end{thebibliography}
\end{document}